    \documentclass{aa}  

\usepackage{txfonts}
\usepackage{graphicx,lipsum}
\usepackage[percent]{overpic}
\usepackage{amssymb,amsmath}
\usepackage[version=4]{mhchem}
\usepackage{newtxtext}
\usepackage[varvw]{newtxmath}
\usepackage{xcolor}

\newcommand{\ignore}[1]{}

\def\aap {{A\&A}}

\def\apjl {{ApJL}}

\def\apjs {{ApJS}}

\def\micron{$\mu$m }
\begin{document}

   \title{ Early planet formation in embedded protostellar disks }
   \titlerunning{ Early planet formation in embedded protostellar disks }

    \subtitle{Setting the stage for the first generation of planetesimals}

   \author{Alex J. Cridland\inst{1}\thanks{cridland@mpe.mpg.de}, Giovanni P. Rosotti\inst{2,3}, Beno\^it Tabone\inst{3}, {\L}ukasz Tychoniec\inst{4}, Melissa McClure\inst{3}, Pooneh Nazari\inst{3}, Ewine F. van Dishoeck\inst{1,3}
          }
    \authorrunning{ Cridland et al. }

   \institute{
   $^1$Max-Planck-Institut f\"ur Extraterrestrishe Physik, Gie{\ss}enbachstrasse 1, 85748 Garching, Germany \\
   $^2$School of Physics and Astronomy, University of Leicester, Leicester LE1 7RH, UK\\
   $^3$Leiden Observatory, Leiden University, 2300 RA Leiden, the Netherlands\\
   $^4$European Southern Observatory, Karl-Schwarzschild-Str. 2, 85748 Garching, Germany
             }

   \date{Received \today}


  \abstract
  {
Recent surveys of young star formation regions have shown that the average Class II object does not have enough dust mass to make the cores of giant planets. Younger Class 0/I objects have enough dust in their embedded disk, which begs the questions: can the first steps of planet formation occur in these younger systems? The first step is building the first planetesimals, generally believed to be the product of the streaming instability. Hence the question can be restated: are the physical conditions of embedded disks conducive to the growth of the streaming instability? The streaming instability requires moderately coupled dust grains and a dust-to-gas mass ratio near unity. Here we model the collapse of a `dusty' proto-stellar cloud to show that if there is sufficient drift between the falling gas and dust, regions of the embedded disk can become sufficiently enhanced in dust to drive the streaming instability. We include four models, three with different dust grain sizes and one with a different initial cloud angular momentum to test a variety of collapse trajectories. We find a `sweet spot' for planetesimal formation for grain sizes of a few 10s of micron since they fall sufficiently fast relative to the gas to build a high dust-to-gas ratio along the disk midplane, but have slow enough radial drift speeds in the embedded disk to maintain the high dust-to-gas ratio. Unlike the gas, which is held in hydrostatic equilibrium for a time due to gas pressure, the dust can begin collapsing from all radii at a much earlier time. The dust mass flux can thus be higher in Class 0/I systems than the gas flux, which builds an embedded dusty disk with a global dust-to-gas mass ratio that exceeds the ISM ratio by at least an order of magnitude. The streaming instability can produce at least between 7-35 M$_\oplus$ of planetesimals in the Class 0/I phase of our smooth embedded disks, depending on the size of the falling dust grains. This mass is sufficient to build the core of the first giant planet in the system, and could be further enhanced by dust traps and/or pebble growth. This first generation of planetesimals could represent the first step in planet formation, and occurs earlier in the lifetime of the young star than is traditionally thought.
}


    \maketitle
%

\section{Introduction}

Some form of planet formation undoubtedly begins early in the life cycle of stars. Such a statement is largely based on the fact that protoplanetary disks - the source of gas for giant planets - tend to last no longer than about 10 Myr before dissipating \citep{Haisch2001}. Moreover, recent high resolution surveys of the dust in protoplanetary disks \citep[][]{vanderMarel2013,Zhang2016,Isella2016,Fedele2018,vT2018,DSHARP1,Andrews2020,VANDAM3} have shown a ubiquity of substructure in extended disks often attributed to the presence of massive planets. This includes substructure in young Class I objects GY 91 \citep{Sheehan2018}, IRS 63 \citep[][]{SC2020} and the well known HL Tau disk \citep{ALMA15}, all of which have ages no older than 500 kyr. Furthermore the few known planets embedded in disks that have been directly observed \citep[i.e.][]{PDS70a,Haffert2019} are relatively massive planets and show signs of effectively being at the end stage of their formation. This is because their inferred accretion rate is very low, and unlikely to contribute much more to their total mass\footnote{The PDS 70 planets have masses of at least 4 M$_{\rm Jup}$ with an accretion rate on the order of $10^{-2}$ M$_{\rm Jup}$Myr$^{-1}$ \citep{Haffert2019}. Even with another 5 Myr (i.e. twice the current age of the system) the planets will gain no more than another 1\% of their current mass.}. 

This observational evidence is suggestive that the bulk of the planet formation process does not occur during the `protoplanetary' disk phase (typically called Class II disks), \ignore{but in an earlier phase when the disk is still surrounded by an envelope of gas and dust - typically called Class I systems. There is strong evidence that significant grain growth is already occurring in the inner region of Class I disks \citep{Harsono2018}, which represents the first major step towards the growth of planets. This, coupled with the aforementioned observations of sub-structure in the continuum emission of the Class I disk IRS 63 \citep{SC2020}, points toward an early start of giant planet formation in an earlier phase of proto-stellar evolution than conventionally thought.} but in an earlier phase when the disk is still surrounded by its nascent envelope of gas and dust - its protostellar stage (Class 0/I). In the protostellar stage the mass flux in the system is split between gas and dust falling onto the embedded disk from the envelope, and mass accretion onto the star through the disk \citep{Hueso2005}. The Class 0 and I are either differentiated by their observed properties such as bolometric temperature and submillimeter excess \citep{Andre1993,Chen1995} or by physical evolution with Class 0 having most of the mass contained in the envelope while Class I systems contain the majority of their mass in the star and embedded disk \citep{vanKempen2009}.

Recent surveys by \cite{Tobin2020} and \cite{Tychoniec2020} studied the distribution of dust masses in Class 0, I, and II objects in Orion and Perseus with ALMA and the VLA and compared them to inferred exoplanet metallicities from \cite{Thorngren2018}. They find that the average Class II disk does not contain enough total dust mass to reproduce the observed population of giant exoplanets. Class 0 and I objects, on the other hand, do contain enough dust mass in their embedded disks to build a giant planet, and require a rather realistic formation efficiency ranging between $\sim 10$ and $30\%$ to do so. Their work demonstrates that, at the very least, physical processes in the Class 0/I phase \citep[at evolution times $<$ 150 kyr,][]{Dunham2014} of proto-stellar evolution lead to the disappearance or conversion of dust mass prior to the Class II phase (ages $> 500$ kyr). In this work we investigate whether planetesimal formation, which represents the first step in the bottom-up planet formation model known as `core accretion', can contribute to the observed loss of dust mass between Class 0, Class I, and Class II objects.

Core accretion broadly describes the growth of a giant planet by first accumulating up to a few 10s of Earth masses (M$_\oplus$) of solid material before then accreting a significant amount of gas until it reaches its final mass of $\sim$Jupiter mass ($=$317 M$_\oplus$) or more. This process differs from the other popular model of planet formation - gravitational instability (GI) that could also be active in the embedded disk during the collapse phase of star formation. The general assumption of GI is that the resulting giant planet would reflect the metallicity of its natal disk and would not generate a significant solid core, making it incompatible with observed planetary populations \citep[as seen in][]{Thorngren2018}. However several numerical studies have shown that through gravitational and/or hydrodynamic interactions between the solid component of the disk and the gravitationally unstable clumps, higher metallicity planets can be generated, possibly including a solid core \citep{Helled2006,Boley2010,Boley2011}. Thus while we recognise that embedded giant planets could have been generated through GI, we focus this work on the production of planetesimals as they possibly represent the seeds of giant planets and certainly represent the seeds of asteroids and comets.

The preferred scenario of building planetesimals from dust grains is known as the streaming instability. This instability arises in regions of a disk where the dust is sufficiently dense, relative to the gas, to have a dynamical effect on the gas flow. This effect locally increase the gas flow velocity, reducing the radial drift speed of the dust. The instability is then fed with more dust from regions where the radial drift rate remains faster, further increasing the dust density in the unstable region. For strictly radial perturbations however, the instability is be suppressed by strong thermal pressure gradients. Thus perturbations in either the vertical or azimuthal direction are also needed for the instability to grow. These perturbations can lead to related instabilities such as the `settling instability' described in \cite{Squire2018,Squire2020}. This build up of dust density further feeds the instability, which causes a runaway process where the dust density rapidly increases until it becomes gravitationally unstable, collapsing into asteroid-sized, bound objects.

The pioneering work on the streaming instability is presented in \cite{YG05}, which showed that the optimal environment for the streaming instability to occur was for dust grains that are moderately coupled to the gas. These types of grains are characterised by a `Stokes number' ($St$) that is about 0.1. The Stokes number is a ratio of a dust grain's `stopping time' to the relevant dynamic timescale of the system, meaning that a grain with a very small Stokes number effectively follows the flow of the surrounding gas, while a larger Stokes number feels a drag force from the gas, but mainly evolves following other external forces like gravity. In addition, the streaming instability requires a relatively high midplane dust-to-gas mass ratio, reportedly between 0.2 and 3 for dust grains with Stokes numbers of $\sim 0.1$. Obtaining such a high dust-to-gas ratio is the main limitation of the streaming instability given that the standard interstellar medium (ISM) ratio is 0.01. In what follows we will largely follow the prescription of \cite{YG05}, with small changes discussed below.

Recent numerical work by \cite{LY2021} has updated the estimates for optimal dust-to-gas mass ratios for a wide range of Stokes numbers. It is our intention to compute the collapse of a `dusty' molecular cloud, which is initialised with the typical ISM dust-to-gas mass ratio, $\epsilon_{\rm ISM}=0.01$, to show that there are regions in the young, embedded disk where the mass ratio exceeds the critical dust-to-gas ratio in order to drive the instability. In regions of the disk where the threshold dust-to-gas ratio can be reached, the streaming instability could turn the available dust mass into planetesimals. 

An early attempt at modelling the `dusty' collapse of a proto-stellar cloud was done by \cite{Morfill1978} who pointed out that ``the calculation of the evolution of the `solid phase' in the collapsing cloud is the only means through which not only the comparison with celestial mechanics information is made possible, but through which the results of meteorite studies can be utilised as well.'' This work can thus help to understand planet formation as a whole, but also help to understand the formation and chemical properties of comets and meteorites. \cite{Morfill1978} find that the dust does indeed fall rapidly, forming a dusty embedded disk with enough dust sedimentation that possible ``dust instabilities will form the seeds for eventual planet formation, and that the planets are formed long before the Sun is formed.'' While this final point may be an over exaggeration of how quickly planets can form in proto-stellar systems, the idea that planet formation should occur \textit{while} the young star is still in its embedded phase has not received much attention since this pioneering work.

\ignore{
More recent numerical studies of a dusty collapse including \cite{BL2017} and \cite{Lebreuilly2020} have demonstrated that dust grains larger than about 10$\mu$m decouple from the gas sufficiently to dramatically alter their collapse trajectory. In particular \cite{BL2017} report a hierarchy of dust dynamics dependent on the grain size. The smallest grains are fully coupled to the gas and maintain nearly the same dust-to-gas ratio throughout the collapse. Grains around 100$\mu$m are sufficiently decoupled that they do fall faster than the pressure-supported gas, but collect along the midplane of the system to form the first embedded dusty disk. Their largest grains (1 mm) are nearly completely decoupled and rapidly fall to the midplane. Near the midplane, these large grains rapidly oscillate around $z = 0$, causing maximum dust-to-gas ratios that exceeds $10^5$ for a short time during the embedded phase. \cite{Lebreuilly2020} similarly show that mm-sized grains produce the most dust enhancement in their magnetised collapse simulations. They additionally find that magnetised models increased the amount of decoupling between gas and dust relative to purely hydrodynamic collapse models because of the extra magnetic pressure supplied to the gas. This extra source of pressure, however, does not drastically alter the grain size-dependent trajectories found in purely hydrodynamic simulations.
}

A recent study of planetesimal formation in a young stellar system by \cite{Drazkowska2018} assumed that the dust in the envelope was strictly monomer-sized (on the order of 0.1 $\mu$m) and that the growth and subsequent enhancement of dust density was driven exclusively by processes in the embedded disk. Their method of enhancing the dust-to-gas ratio relied on the well known traffic jam effect \citep{Pinilla2016} at the water ice line of the embedded disk. This effect drives large dust-to-gas ratios because the removal of the water ice layer from the grains makes them more susceptible to fragmentation. The maximum grain size is thus reduced by an order of magnitude \citep[as seen in][]{B12}, reducing the radial drift rate by a corresponding amount. Hence the radial drift rate leaving the ice line is an order of magnitude lower than the approaching rate which drives a density enhancement that is ideal for the streaming instability. \cite{Drazkowska2018} find that only a minimal amount of their final distribution of planetesimals were formed during the Class I phase of their model, with the majority still being formed in the Class II phase. 

The assumption of strictly monomer size dust grains in the envelope is not supported by continuum observations of young stellar systems. Preliminary evidence in the youngest dense cores already show evidence of grain growth up to $\sim 1$\micron sized grains through infrared `core-shine' \citep{Pagani2010}; a process where background light is strongly scattered by larger grains in the core. In the slightly more evolved pre-stellar core, L1544, \cite{CT2019} found changes in the millimetre opacity towards the inner $\sim 2000$ AU indicative of grain sizes up to $3-4~\mu$m. Surveys of Class 0 envelopes have found low dust emissivity indices that are consistent with mm-sized dust grains \citep{Jorgensen2009,Kwon2009,Galametz2019}. Class I objects also show similarly small emissivity indices that indicate dust grain growth up to between 100 \micron and mm-sized grains \citep{AgurtoG2019}. Furthermore, modelling efforts at sub-millimeter wavelengths require maximum grain sizes up to $\sim 1$mm to explain observed continuum emission profiles \citep{Miotello2014}, and polarized dust emission studies have shown evidence for grain growth up to at least 10 \micron during the early stages of the collapse \citep[$t\sim 10^5$ years,][]{Valdivia2019}. 

Theoretical models can account for some of this grain growth within the lifetime of the collapse. For example \cite{Guillet2020} report hydrodynamical turbulence-driven grain growth of up to 1-10 \micron in one free-fall time of their collapse simulation. \cite{Ormel2009} argue that dust growth, up to a size of 100 $\mu$m, is only possible if additional support mechanisms like ambipolar diffusion delay the collapse to longer than a free-fall time, or if a significant ice mantle is present on the grains. Otherwise, dust growth is inefficient during the collapse phase of their models. Given the typically cold temperatures of the outer envelopes of pre- and proto-stellar objects and the fact that water ice makes up a large fraction of the overall oxygen abundance \citep{Boogert2015,vanDishoeck2021} the latter of the above two requirements is very likely. Even with this requirement met, however, whether grain growth within the envelope can produce the observed $\sim$mm grain sizes is still an open question.

\ignore{
Instead, observational studies of envelopes around proto-stars have shown a variety of possible maximum grain sizes, exceeding the monomer size by at least an order of magnitude. \cite{Galametz2019} found low emissivities indicative of mm-sized dust grains (four orders of magnitude above the monomer size) in many Class 0 envelopes. Meanwhile \cite{AgurtoG2019} studied the inner envelope of the Class I object Per-emb-50 and found that the maximum grain size was limited to 100 \micron by their observations. For younger systems, \cite{CT2019} found evidence for a change in opacity in the inner $\sim 2000$ AU of L1544 which suggested dust grain sizes up to between 3-4 \micron were present in that region. The important distinction between envelopes containing only monomer sized grains, and those that contain larger dust grains is that the collapse trajectory of large dust grains will diverge with respect to the collapse of monomer-sized grains because of differences in their gas coupling. Larger grains are less coupled to the gas and do not directly feel the effect of the gas pressure, but feel an aerodynamic drag as they fall towards the centre of the cloud. They hence fall faster than the gas, losing their angular momentum to drag forces in the process. This loss of angular momentum should let them reach smaller radii than the gas when they intersect the midplane of the embedded disk. The different collapse trajectory between the dust and gas could then lead to enhancements in the dust-to-gas ratio that is ideal for the streaming instability. 
}

More detailed hydrodynamical studies of a dusty collapse including \cite{BL2017} and \cite{Lebreuilly2020} have demonstrated that dust grains larger than about 10$\mu$m decouple from the gas sufficiently to dramatically alter their collapse trajectory. In particular \cite{BL2017} report a hierarchy of dust dynamics dependent on the grain size. Monomer-sized grains are fully coupled to the gas and maintain nearly the same dust-to-gas ratio throughout the collapse. Grains around 100$\mu$m are sufficiently decoupled that they do fall faster than the pressure-supported gas, and collect along the midplane of the system to form the first embedded dusty disk. Their largest grains (1 mm) are nearly completely decoupled and rapidly fall to the midplane. Near the midplane, these large grains rapidly oscillate around $z = 0$, causing maximum dust-to-gas ratios that exceeds $10^5$ for a short time during the embedded phase. \cite{Lebreuilly2020} similarly show that mm-sized grains produce the most dust enhancement in their magnetised collapse simulations. They additionally find that magnetised models increased the amount of decoupling between gas and dust relative to purely hydrodynamic collapse models because of the extra magnetic pressure supplied to the gas. This extra source of pressure, however, does not drastically alter the grain size-dependent trajectories found in purely hydrodynamic simulations.

These simulations demonstrate the key feature that we wish to test with our semi-analytical model presented here: that differences in the gas and dust trajectories caused by the decoupling of large grains lead to a build up of dust density sufficiently large to induce the streaming instability - possibly leading to the generation of the first generation of planetesimals. Given the simplicity of our semi-analytic approach we can extend the above studies by connecting our results directly to the physics of planetesimal formation and the global mass evolution of embedded disks. Our simulations focus on a single grain size, thereby ignoring the impact of dust grain growth during the collapse. We test a range of dust grain sizes to probe what effect different levels of hydrodynamic coupling have on the clumping and future collapse due to the streaming instability. The paper is organised as follows: in section \ref{sec:method} we introduce our computational method and outline the physics on which we focus. In section \ref{sec:results} we present the results of our collapse calculations, including estimates of the evolution of the local dust-to-gas ratio. In section \ref{sec:discuss} we discuss the relevance of our results and estimate the mass conversion of dust mass to planetesimals while in section \ref{sec:conclude} we present our conclusions.

\section{Methods}\label{sec:method}

\begin{figure*}
    \centering
    \includegraphics[width=\textwidth]{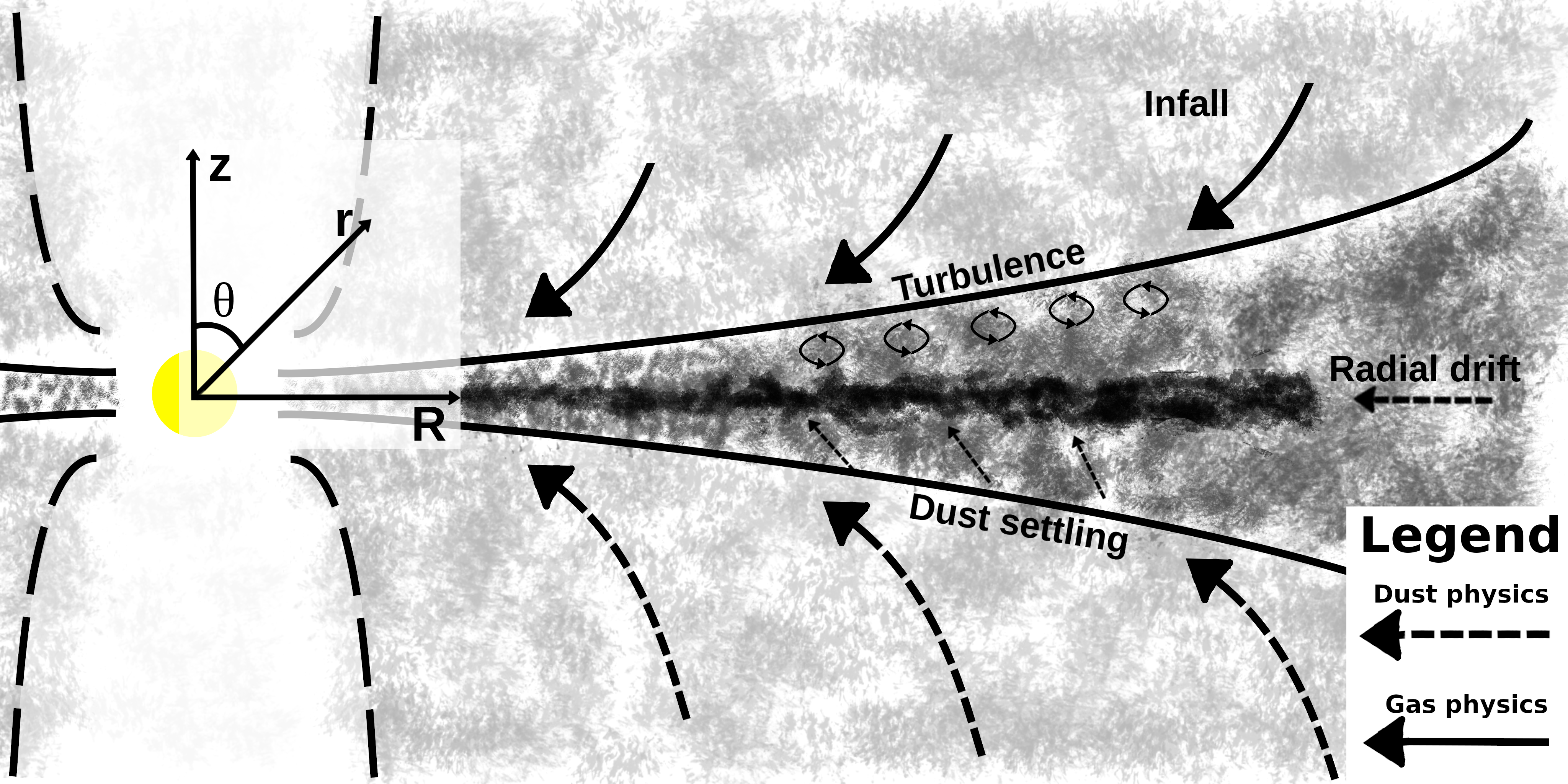}
    \caption{A schematic view of the embedded disk within a collapsing envelope. Unlike in traditional `protoplanetary disks' - i.e. Class II objects - embedded disks (Class 0/I) are continually fed with new gas and dust from the envelope (infall). The gas in the embedded disk is subject to turbulent flow and evolves viscously. The dust, on the other hand, tends to sink to the midplane of the embedded disk and radially drifts towards the host star. Since the gas, but not the dust, is held aloft by gas pressure, the gas and dust arriving at the embedded disk \textit{at any given time} originated from different initial locations in the envelope, with dust arriving from larger initial radii than the gas. The grey scale broadly shows the total volume mass density of the gas and dust in the proto-stellar system.}
    \label{fig:env_collapse}
\end{figure*}

The key physics that we wish to capture with our model is shown in Figure \ref{fig:env_collapse}. The gas and dust feel different forces during their infall phase which changes the way that they collapse towards the midplane. While the gas is largely held in place by a hydrostatic balance between gravity and gas pressure (until the arrival of the rarefaction wave), the dust feels no such gas pressure. Dust grains begin their collapse immediately while the gas must wait for the loss of gas pressure. 

While the dust moves through the gas, however, it feels a headwind that drains angular momentum and alters its trajectory with respect to pure projectile motion. This effect is especially felt when a dust grain enters the embedded (rotationally supported) gas disk, where it quickly begins to radially drift through the disk. Dust mass is thus removed from the midplane of the embedded disk - the optimal location for the generation of planetesimals - and is replenished by incoming dust that accretes onto, then settles down from, the upper surface of the embedded disk.

\subsection{Gas model}\label{sec:gasmethod}

The backbone of this work is the semi-analytic collapse model of \cite{Visser2009}. Here we outline some of the primary features of the model that are important for our current work. The model computes the collapse of an initially isothermal, spherical, and axisymmetric cloud of gas following the theoretical work of \cite{Shu1977} and \cite{TSC1984}. The slow rotation and angular momentum conservation forces the infalling gas towards the equatorial plane, forming an embedded disk. Our model follows the work of \cite{CM1981} in handling the flux of gas onto the central star and the disk. The mass flux onto the disk modifies the usual diffusion equation to have the form \citep{LB74,Dullemond2006}:\begin{align}
    \frac{\partial\Sigma}{\partial t} &= -\frac{1}{R}\frac{\partial}{\partial R} \left(\Sigma R u_R\right) + S,
    \label{eq:hydro}
\end{align} 
where $\Sigma$ is the disk gas surface density, and $S = 2 N\rho_S u_z$ is the mass flux from the envelope through both the bottom and top surfaces of the disk - hence the factor of 2. $\rho_S$ is the volume mass density of the gas at the disk-envelope interface and $N$ is a normalisation factor that ensures an equilibrium between the mass accretion rates onto the disk, through the disk, and onto the central star. In the \cite{Visser2009} model the interface between the disk and envelope is defined where the gas pressures in the disk and envelope are equal. The radial speed of the gas while in the disk, in the direction of the cylindrical radius $R$, is:\begin{align}
    u_R(R,t) = -\frac{3}{\Sigma\sqrt{R}}\frac{\partial}{\partial R} \left( \Sigma \nu\sqrt{R}\right),
    \label{eq:speed}
\end{align}
where \begin{align}
    \nu(R,t) = \alpha c_{s,d} H
\end{align}
is the viscosity under the standard $\alpha$-prescription \citep{SS73}, with $\alpha=0.01$. Equations \ref{eq:hydro} and \ref{eq:speed} are only valid if the specific angular momentum of the inflowing gas is the same as in the disk where it is delivered. This is not always true \citep{LinPringle1990} and our model contains an extra term in equation \ref{eq:speed} to account for the re-distribution of angular momentum of incoming gas based on \cite{Hueso2005}. Equation \ref{eq:speed} thus becomes: \begin{align}
    u_R(R,t) = -\frac{3}{\Sigma\sqrt{R}}\frac{\partial}{\partial R} \left( \Sigma \nu\sqrt{R}\right) - \eta_r\sqrt{\frac{GM}{R}},
    \label{eq:speed2}
\end{align}
where $\eta_r = 0.002$ is a numerical factor that best reproduces the results of \cite{Hueso2005}.

The disk gas scale height $H$ and disk sound speed $c_{s,d}$ are linked through the Kepler rotation rate:\begin{align}
    \Omega_k(R,t) = \sqrt{\frac{GM(R,t)}{R^3}},
\end{align}
where $M(R,t)$ is the total mass inward of $R$ at time $t$ including the host star such that:
\begin{align}
    H(R,t) = \frac{c_{s,d}}{\Omega_k}.
\end{align}

The cloud initially has a solid body rotation rate of $\Omega_0$ and an initial gas volume density that varies with the spherical radius $r$ \citep{Shu1977}:\begin{align}
    \rho_0(r) = \frac{c_s^2}{2\pi G r^2},
    \label{eq:initial}
\end{align}
where $G$ is the gravitational constant, and $c_s$ is the effective gas sound speed of the cloud. We set a maximum envelope radius $r_{\rm env}$, so that the initial cloud mass is:\begin{align}
    M_0 \equiv \frac{2 c_s r_{\rm env}}{G} = 1 M_\odot.
\end{align}
The structure and evolution of the volume density of the gas depends on the spherical radius $r$, polar angle $\theta$ and time $t$ through the dimensionless functions $\mathcal{A}$, $y$, and $w$ \citep{CM1981,TSC1984}:\begin{align}
    \rho(r,\theta,t) &= \frac{\mathcal{A}(x,\theta,\tau)}{4\pi G t^2} \\
    u_r(r,\theta,t) &= c_s y(x,\theta,\tau) \\
    u_\theta(r,\theta,t) &= c_s w(x,\theta,\tau)
\end{align}
and dimensionless variables $x = r/c_s t$ and $\tau = \Omega_0 t$. The rotational speed $u_\phi$ is kept constant in this solution, assumes solid body rotation at a rate of $\Omega_0$. The values of $\mathcal{A}$, $y$, and $w$ are derived numerically in \cite{TSC1984}\footnote{In \cite{TSC1984} the function $y$ is represented with a $v$. To avoid confusion with the dust velocity shown below we have changed our notation.}. The cylindrical radius $R$, height above the disk midplane $z$, spherical radius $r$, and the polar angle $\theta$ are related through:\begin{align}
    R &= r\sin\left(\theta\right)\nonumber\\
    z &= r\cos\left(\theta\right).\nonumber
\end{align}

Broadly speaking the cloud collapses inside out, with a rarefaction wave (or collapse front) moving outward at the sound speed \citep{Shu1977}. The wave front has the dimensionless position of $x=1$ over all time. At a given location with $x^\prime>1$, the gas remains in hydrostatic balance between its gravity and pressure from all gas with $x < x^\prime$. Once $x^\prime=1$ the shell loses its supporting gas pressure and begins to collapse towards the centre of the cloud. Due to a Taylor expansion in their derivation, when $x^\prime=\tau^2$ the \cite{TSC1984} solution breaks down and the model switches to use the solution derived by \cite{CM1981}.

The \cite{CM1981} solution follows the trajectory of individual gas streamlines from the envelope to the disk rather than building a self-similar solution like was done in \cite{Shu1977} and \cite{TSC1984}. The evolution of the gas density, radial, polar, and rotational velocities are:\begin{align}
    \rho(r,\theta,t) &= -\frac{\dot{M}}{4\pi r^2 u_R}\left[1 + 2\frac{R_c}{r} P_2(\cos\theta_0)\right]^{-1},\\
    u_r(r,\theta,t) &= -\sqrt{\frac{G M}{r}}\sqrt{1 + \frac{\cos\theta}{\cos\theta_0}},\\
    u_\theta(r,\theta,t) &= \sqrt{\frac{G M}{r}}\sqrt{1 + \frac{\cos\theta}{\cos\theta_0}}\frac{\cos\theta_0 - \cos\theta}{\sin\theta},\\
    u_\phi(r,\theta,t) &= \sqrt{\frac{G M}{r}}\sqrt{1 + \frac{\cos\theta}{\cos\theta_0}}\frac{\sin\theta_0}{\sin\theta},
    \label{eq:CM81}
\end{align}
respectively, where $R_c=16^{-1}c_sm_0^3t^3\Omega_0$ is the centrifugal radius, $\dot{M} = m_0c_s^3/G$ is the total accretion rate from the envelope onto the host star and disk, $m_0 = 0.975$ is a numerical factor, $P_2$ is the second order Legendre polynomial, and $\theta_0$ is the polar angle with which a given streamline enters into the \cite{CM1981} solution. The outer shell of the envelope reaches the star and disk at a time of $t_{\rm acc} = M_0/\dot{M}$ ($=252.5$ kyr in our model) and signifies the end of the collapse phase of the proto-star. \cite{Visser2009} points out, however, that this end point does not exactly corresponds to other `mature' Class II systems, as the disks produced here tend to be warmer at $t=t_{\rm acc}$ than the average protoplanetary disk observed in the literature.

The gas collapse is computed for two different envelope initial conditions: the so called `reference' case with $\Omega_0 = \Omega_{\rm ref} \equiv 10^{-13} \rm s^{-1}$ and the `standard' case with $\Omega_0 = \Omega_{\rm std} \equiv 10^{-14} \rm s^{-1}$ \citep[cases 7 and 3 respectively from][]{Visser2009}. Both simulations are run for the same initial cloud mass $M_0 = M_\odot$ and same initial envelope size $r_{\rm env} = 6686~ \rm AU$ which corresponds to a gas sound speed of $0.26~\rm kms^{-1}$ and a gas temperature of approximately $18~\rm K$. The main difference between these simulations is that the resulting embedded disk is larger, in which case because of the dependence of the centrifugal radius $R_c$ on $\Omega_0$.

\subsection{Tracing individual dust trajectories}

Included in the model is a post-processing scheme to trace the trajectories of Lagrangian particles throughout the collapse. This method has been used to trace the collapse of individual parcels of gas in order to study their chemical evolution \citep{Visser2009,Visser2011,Harsono2013,Drozdovskaya2016}. The method was designed to sample the local gas density and temperature, the local gas velocity, and the dust temperature to determine the properties relevant to chemistry as well as the direction that the parcel will next evolve. Here we use the fact that the parcel algorithm looks up the local gas density and velocity, and modify it to compute the collapse trajectory of dust in the envelope, which we outline below.

The primary difference between the trajectory of the dust and gas is that dust with sufficiently high Stokes number does not feel any direct impact of the gas pressure. As such, dust parcels that are only moderately coupled to the gas - with Stokes numbers on the order of 0.1-1 - begin their collapse at $t=0$ rather than when $x=1$. Because it takes a finite time for the collapse front to reach the other regions of the envelope, there is an immediate drift between the collapse trajectories of the gas and dust there. As the dust accelerates towards the centre of the cloud, differences between the gas and dust velocities cause drag that drains angular momentum from the dust, accelerating their collapse towards the equatorial plane and producing a dusty disk with a smaller outer radius than the gas disk.

The drift felt by the dust is proportional to the total velocity difference between the gas and dust \citep{W77}. In the Epstein regime, when the mean free path of the gas is longer than the size of the dust grain ($s$), the hydrodynamical coupling of the dust grains, known as its stopping time is defined as \citep{Whip73}:\begin{align}
    t_s = \frac{\rho_s s}{\rho c_s},
\end{align}
where $\rho_s$ is the density of an individual grain (assumed here to be 3.6 $\rm g~cm^{-3}$) and $\rho$ is the volume density of the surrounding gas. The Stokes number of the grain is the dimensionless quantity of the stopping time, defined as $St = \Omega t_s$. In the envelope, we assume that $\Omega = \Omega_0$, the solid body rotation rate of the cloud, while in the disk $\Omega=\Omega_K$, the Kepler rotation frequency. When the mean free path of the gas is nearly the size of the dust grain the drag on the dust enters into the Stokes regime, however nowhere in the collapse does this occur. In the Epstein regime, the drag force is \citep{W77}:\begin{align}
    \vec{F}_D = \frac{4\pi}{3}\rho s^2 c_s \vec{\Delta v},
\end{align}
where $\vec{\Delta v} = \vec{u} - \vec{v}$ is the difference between the gas ($\vec{u}$, computed in section \ref{sec:gasmethod}) and dust ($\vec{v}$) velocities. The acceleration of the dust due to gas drag is thus:\begin{align}
    m\vec{a}_D = \frac{4\pi}{3}\rho_s s^3 \vec{a}_D &= \frac{4\pi}{3}\rho s^2 c_s \vec{\Delta v}\nonumber\\
    \vec{a}_D &= \left(\frac{\rho}{\rho_s}\right)\frac{c_s\vec{\Delta v}}{s} = \frac{\vec{\Delta v}}{t_s}.
    \label{eq:accdrift}
\end{align}
When the dust is falling faster than the gas ($\Delta v < 0$) it feels a headwind that slows its descent, while in the opposite case ($\Delta v > 0$) the dust is swept up by the falling gas. 

At the beginning of the trajectory calculation $10^5$ dust parcels are placed randomly throughout the cloud with spherical radii between $100~{\rm AU}<r<r_{\rm env}$ and polar angles between $0<\theta<\pi/2$. Initially each dust parcel begins with $v_R = v_z = 0$ and $v_\phi \equiv \dot{\theta} = \Omega_0 (\rm s^{-1})$. These parcels begin falling immediately since they are not held up by hydrostatic balance. As such, dust parcels whose initial position can be described with $x > 1$ begins to collapse before the gas that surrounds them. This setup at $t = 0$ assumes that the dust and gas are both distributed homogeneously through the core prior to the beginning of the collapse. Given the differential collapse that we expect to see, having both species starting with the same spatial distribution is likely too simple. In reality the spatial distribution of the gas and dust will be set by turbulent motions in the larger molecular cloud, and may largely begin with a more compact dust distribution than the gas. Such a distribution would thus lead to an enhanced differential collapse between the gas and dust than we will simulate here.

The equations of motion for the dust parcels are:\begin{align}
    a_R = \frac{d v_R}{dt} &= -\frac{GM(r)}{r}\frac{R}{r} + R v_\phi^2  + a_{D,R}\qquad &(\rm cm~s^{-2}) \label{eq:main1}\\
    a_\phi = \frac{d v_\phi}{dt} &= - \frac{2 v_R  v_\phi}{R} + a_{D,\phi}\qquad & (\rm s^{-2}) \label{eq:main2}\\
    a_z = \frac{d v_z}{dt} &= -\frac{GM(r)}{r}\frac{z}{r} + a_{D,z}\qquad & (\rm cm~s^{-2})
    \label{eq:main3}
\end{align}
where $r$ is the magnitude of the spherical radius. We note the units of each equation above to recognise that $v_R$ and $v_z$ are in linear units, while $v_\phi$ is the rotational speed in angular units. While the simulation is itself in two dimensions ($R$,$z$), gas drag depends on the difference between the gas and dust velocities in all three dimensions. Rather than change the underlying code of \cite{Visser2009} to include a third dimension, we assume that the coordinate system rotates with the rotating dust parcel at $v_\phi$. As such $\phi = 0$ for all time in the rotating coordinate system, and we include a Coriolis force (third term in eq. \ref{eq:main2}). The mass contributing to the gravitational acceleration, $M(r)$, includes the cloud and disk mass orbiting inward of the dust parcel, as well as the host star. 

We evolve the velocity over every time step following the explicit expression of \cite{Rosotti2016}. The velocity in the next timesteps follows:\begin{align}
    \vec{v}(t + \Delta t) &= \vec{v}(t)\exp\left(-\Delta t/t_s\right) + \vec{a}_{\rm gas}\Delta t +\nonumber\\
    & \left[ \vec{u}(t) + \left(\vec{a}_{\rm dust}-\vec{a}_{\rm gas}\right) t_s \right]\left(1 - \exp\left(-\Delta t/t_s\right)\right).
    \label{eq:evolve}
\end{align}
Expressing the velocity evolution in this way ensures that in the event that the timestep exceeds the stopping time of the grain, the algorithm remains well behaved. In this limit ($\Delta t>>t_s$) the dust parcel `forgets' its previous velocity and evolves with the gas, but with a slight drift proportional to the difference between the gas and dust accelerations and the stopping time. In the opposite limit ($\Delta t << t_s$) there is insufficient time for the dust drag to alter the trajectory of the dust, and it evolves as if it is in free-fall. 

The gas evolution, including the gas velocities ($\vec{u}$), are pre-computed using the model of \cite{Visser2009}, hence we do not self-consistently evolve the gas using $\vec{a}_{\rm gas}$, nor do we include any back-reaction on the gas evolution caused by the clumping of dust. For the purposes of equation \ref{eq:evolve}, however, we do compute the forces on the gas at the instantaneous position of the dust parcel. Before the collapse wave reaches an annulus of gas with $x>1$ we assume that $a_{\rm gas}=0$, $u_R = u_z = 0$, and $u_\phi = \Omega_0$. When the dust parcel reaches an annulus of gas with $x<1$ we compute the acceleration felt by the gas:\begin{align}
    \vec{a}_{\rm gas} = \left(-\frac{GM(r)}{r} - \frac{\nabla P}{\rho} \right)\vec{\hat{r}} + Ru_\phi^2 \hat{R} - \frac{2 u_Ru_\phi}{R} \hat{\phi},
    \label{eq:gasevolve}
\end{align}
where $\hat{R}$ and $\hat{\phi}$ are unit vectors in the direction of the cylindrical radius and rotation, and $\vec{\hat{r}}$ is the unit vector in the spherical radius direction. The pressure gradient $\nabla P$ and local gas density $\rho$ are taken from the pre-computed gas properties of the \cite{Visser2009} collapse model.

Equations \ref{eq:main1}-\ref{eq:gasevolve} are implemented into a routine that is called by the DVODE integrator \citep{dvode} to integrate the trajectories of the collapsing dust particles. Before the dust trajectories are computed, we compute the evolution of the gas collapse using the methods described in section \ref{sec:gasmethod}. With these pre-computed collapse models we then trace the trajectories of each dust parcel using the method described in this subsection. The routine which computes the dust particle's speed uses the particle's current location to estimate the gas' current speed and density from the pre-computed collapse model. When the dust particle is in the disk, and when the evolution of the surrounding gas is described by the \cite{TSC1984} solution we assume that the gas rotates around the $R=0$ axis with sub-Keplarian and solid body rotation respectively. When in a region where the gas evolution can be described by the \cite{CM1981} solution we assume the gas rotates as is shown in equation \ref{eq:CM81}. The routine uses the gas speeds as inputs for computing $\vec{\Delta v}$ and the current evolution of the dust's speeds. A benefit of using DVODE is its adaptive time stepping which efficiently handles early times when the dust motion is slow, and later times when the dust parcel is radially drifting through the embedded disk, and hence the problem becomes more stiff.

\subsection{Post processing the simulations}

We compute the trajectory of 100,000 dust parcels and with these trajectories we trace the mass evolution of the dusty envelope by `tagging' each dust parcel with a particular amount of initial dust mass and later investigating where that mass is delivered. This is done as a post-processing of the simulation and is outlined here. The general evolution of a dust parcel is to rapidly fall from out of the envelope into the disk after which it begins to radially drift into the central star. The initial collapse rate and later radial drift rate are all dependent on the size of the dust grain, and is discussed in more detail in the following section. For some of the dust parcels, the above algorithm forces very small time steps which ultimately makes their further evolution intractable. This generally occurs for particles entering in the disk at later times when the gas densities and radial drift rates are the highest. For these particles we extrapolate their evolution based on their last successful integrate step and the assumption that they are radially drifting in the disk.

\subsubsection{Trajectory extrapolation}

The extrapolation step takes any dust parcel for which further integration becomes intractable, takes its position $R_{\rm last}$, $z_{\rm last}$ and time $t_{\rm last}$, to estimate and their final position at $t_{\rm acc}$. Since all of the parcels that grind to a halt can be found in the disk, the extrapolation is done using two assumptions: the first is that the velocity in the cylindrical radius direction is `forgotten' after one stopping time has elapsed. The radial speed is then replaced by the typical radial drift speed defined as \citep{B12}:\begin{align}
    v_{\rm drift} = -\frac{2}{St + St^{-1}}\frac{c_s^2}{v_\theta}\gamma,
    \label{eq:drift}
\end{align}
where $\gamma = |d\log P/d\log R|$ is the pressure gradient in the gas, and $v_\theta = \Omega R$ is the rotational speed in the disk (where $\Omega$ is the Kepler frequency). This is done smoothly, so that the extrapolated radial speed is:\begin{align}
    v_R(\Delta t) = v_R(t_{\rm last})\exp{(-\Delta t/t_s)} + (1 - \exp{(-\Delta t/t_s)})v_{\rm drift},
\end{align}
where $v_R(t_{\rm last})$ is the last recorded radial speed that the simulation computed for a particular dust parcel, and $\Delta t$ is the amount of time since $t_{\rm last}$ where $v_R$ is being computed. In this way the extrapolated radial position is:\begin{align}
    R(\Delta t) = R_{\rm last} + v_R \Delta t.
\end{align}

The second assumption refers to the $z$-direction, where a typical particle trajectory in the disk is to rapidly approach the midplane, then proceed into a damped oscillation around $z=0$. In the simulation itself, particles that cross $z = 0$ are mirrored into the $z > 0$ plane and thus these oscillations are not observed from the code's typical output. To account for this oscillatory motion we extrapolate the z-position of the dust parcel as follows:\begin{align}
    z(\Delta t) = \left|\frac{v_z(t_{\rm last})}{\Omega}\sin{\left[(\Delta t\Omega -\phi_{\rm last})\right]}\right|\exp{(-\Delta t\Omega)},
\end{align}
where $\phi_{\rm last}$ ensures that when $\Delta t = 0$, $z = z_{\rm last}$, and $\Omega$ has the same definition as it did in equation \ref{eq:drift}. In this way the extrapolated particles both oscillate and are damped towards the midplane on a dynamical timescale. The damping rate of the dust must also depend on the stopping time of the grain because of the role that turbulent mixing has on slowing dust settling. Hence the smaller grains may be damped slower than the larger grains do. However since the original calculation of the dust collapse ignores the impact of diffusion (discussed below) we will also neglect its effect during the extrapolation.

\ignore{
Both of the above methods for extrapolation assume that the particle is in the disk at the point that their trajectory requires extrapolation. As is shown below, a typical trajectory in our collapse does include the particle reaching the embedded disk prior to its need for extrapolation. Part of the reason that the evolution of some particles grind to a halt is due to the rapid change in gas density in the transition region between the collapsing envelope and the embedded disk.
}

\subsubsection{Dust mass assignment}\label{sec:method_assignment}

To compute the dust-to-gas mass ratio throughout the embedded disk we must track the movement of dust mass throughout the collapse. Each dust parcel is assigned a certain dust mass at the beginning of the collapse. This initial assignment is done as follows: first a logarithmically spaced grid from 0.1 AU out to $10^4$ AU in both the cylindrical radius and height is set at $t=0$. The average gas density in each of these cells is computed from the initial gas density distribution (equation \ref{eq:initial}) and the initial dust density is set to $\rho_{\rm dust,0} = \epsilon_{\rm ISM}\rho_{\rm gas,0}$. The total dust mass in a given grid cell is computed based on the volume\footnote{Each cell has the shape of a thin shell with thickness $dR$ and height $dz$, each computed on our logarithmic grid.} of the cell and its average density. This dust mass is then equally distributed to all dust parcels that are present in the grid cell. We assume that the dust mass in each parcel is conserved during the collapse - so that no diffusion occurs in or out of the parcels. 

In all future timesteps we use the same grid that was defined before and locate all dust parcels that have entered into a given grid cell. We sum their dust mass and compute the average dust density in that cell using this mass and the physical volume of the cell. Finally we use the average gas density in the grid cell, from the \cite{Visser2009} model and the dust density to compute the local dust-to-gas ratio in the cell. This is then repeated for each grid cell. Since a dust parcel must be in a given cell at a given timestep in order to contribute a dust mass, some regions of the disk will show a remarkably low dust-to-gas ratio in cells where no dust parcels are present. We do not claim that regions like this will exist in practice, since diffusion, which we ignore, will play a role in spreading the dust mass more evenly through the disk.

\subsubsection{Defining the available mass for the streaming instability}

\cite{YG05} show that the optimal environment for the streaming instability is dust grains that are moderately coupled to the gas ($St < 0.1$) and with relatively high dust-to-gas ratios. Their optimal dust-to-gas mass ratios were 0.2 and 3 with a slight reduction in growth speed at ratios of exactly unity caused by the instability wave speed reaching a minimum. Recent numerical work by \cite{LY2021} have updated the estimates for optimal dust-to-gas mass ratios for a wide range of Stokes numbers. They report an updated relation between the grain Stokes number and threshold mass density ratio for efficient particle clumping, leading to the streaming instability, of:\begin{align}
    \log_{10}(\epsilon_{\rm crit}) = A(\log_{10}(St))^2 + B\log_{10}(St) + C,
    \label{eq:fit}
\end{align}
where $\epsilon\equiv \rho_d/\rho_g$ is the dust-to-gas ratio \citep{LY2021}. There are two regimes that show different Stokes-dependencies on the mass ratio threshold. Thus their fitted constants have the values $A=B=0$ and $C=\log_{10}(2.5)$ when $St < 0.015$, and $A=0.48$, $B=0.87$, and $C=-0.11$ in the opposite case. In each cell that contains dust particles, we compare the computed dust-to-gas ratio to the threshold in equation \ref{eq:fit} and in any cells where the threshold is reached we assume that that dust is available for the streaming instability.

\subsection{Caveats}
\ignore{ 
In Figures \ref{fig:smdust_gas-to-dust} and \ref{fig:smdust_em13_gas-to-dust}, the bulk of the dust mass is found in only three places: along the midplane, clustered near the surface of the gas disk (grey solid line), or still in the envelope (top right of each panel).
}

One simplification of our model is the lack of turbulent diffusion acting on our collapsing dust parcels. As a result we find that dust settling is very efficient in our model of an embedded disk. For the sake of setting the accretion rate through the embedded disk we assume that the turbulent $\alpha = 0.01$, which would imply a high degree of turbulence. Strong turbulence in the vertical direction would tend to act against dust grain settling - particularly for the smallest grains, which would spread the dust mass off of the midplane more smoothly throughout the whole disk. In the standard case of isotropic turbulence, one generally assumes that the viscosity driving mass accretion through the disk and the turbulence in the vertical direction are described with the same $\alpha$-parameter. If the turbulence is not isotropic, or if the angular momentum transport responsible for disk accretion is driven by another physical process - like disk winds - then the vertical $\alpha$ and the disk viscosity $\alpha$ could be different. 

The planetesimal formation model of \cite{Drazkowska2018} favoured lower turbulence ($\alpha\sim 10^{-4}$) in the vertical direction and intermediate turbulence ($\alpha = 10^{-2}-10^{-3}$) in the radial direction (i.e. responsible for disk accretion). Observational evidence also tends to favour lower vertical turbulence \citep[e.g.][]{Pinte2016} in young proto-stellar systems, while observational constraints on the role of turbulence in global disk accretion have yet to be confidently found. Our method of assuming a high $\alpha$ for the global evolution of the embedded gas disk while ignoring the impact of vertical turbulence on the evolution of the dust is a simplification, but is consistent with a disk model that drives angular momentum transport through MHD disk winds, but otherwise has low intrinsic turbulence. Such a model is consistent with observational and theoretical constrains, but generally evolves differently - both in size and surface density distribution - from what we assume in our model see for example \citep[see for example][]{Tabone2021}.

We assume here that the dust can be described by a single grain size. Young clouds, however, are better described by a distribution of dust grain sizes with a \textit{maximum} grain size that is similar to the ones that we have modelled in this work. Furthermore, the dust mass distribution can change through coagulation and fragmentation which greatly complicates the nature of this problem. Depending on the efficiency of grain growth the mass of the larger grains could be maintained against radial drift for an extended period of time; which changes the landscape that we investigate here. The large grains have the highest Stokes numbers and thus require lower dust-to-gas mass ratio thresholds to drive the streaming instability than the smaller grains (recall equation \ref{eq:fit}). This would allow the contribution of more regions along the midplane of the disk to the production of planetesimals, potentially increases its efficiency. Including a simple formulation of dust grain growth would be an interesting next step that is currently beyond the scope of this work.

Finally, our simulation ignores the impact on magnetic fields on both the collapse trajectory of the gas and the dust. As previously mentioned \cite{Lebreuilly2020} found that models which include magnetic fields cause the drift between collapsing dust and gas to be enhanced, mainly driven by the fact that magnetic fields provide an extra source of pressure support on the gas but not the (neutrally charged) dust. Dust grains can also carry a charge by absorbing free electrons that are liberated from elemental and molecular sources through thermal and radiative effects \citep[explored for ex. in][ for HII regions]{Akimkin2015}. They would thus feel a dynamical impact from the large scale magnetic fields in the envelope during their collapse. Furthermore the magnetic fields offer torques which impact the distribution of the angular momentum of the gas as it collapses. Here, choosing different initial angular momentum models provides a partial accounting of this physical effect because the standard model results in smaller disks, as predicted in magnetised collapses and noted in \cite{Visser2009}. The extra gas pressure found by \cite{Lebreuilly2020} might enhance the build up of dust-to-gas at earlier times because of a further delay in the collapse of the gas.

\section{ Results: Dusty collapse }\label{sec:results}

Below we present the results of four collapse simulations based on the prescriptions above. Three of the simulations use initial conditions consistent with the `standard case' from \cite{Visser2009}: initial solid body rotation rate $\Omega_{\rm std} = 10^{-14} {\rm s}^{-1}$, cloud mass M$_0$ = 1 M$_{\odot}$, and average sound speed $c_{\rm s,0}=$ 0.26 km s$^{-1}$. This sound speed is equivalent to an average gas temperature of 18.5 K. The parameter that varies between these three simulations is the assumed size of the dust grain. These sizes are  0.34 $\mu$m (consistent with monomer sizes, known as the Std-sml model), 34 $\mu$m \citep[consistent with][known as the Std-med model]{AgurtoG2019}, and 3.40 mm \citep[consistent with][known as the Std-lrg model]{Galametz2019}. Their respective sizes are consistent with average Stokes numbers at the beginning of the simulation of 0.0012, 0.12, and 12 respectively. We define the initial Stokes number in the envelope as:\begin{align}
    St_{\rm env} = \frac{\rho_s s\tau_{\rm acc}}{\rho_{\rm g,0}c_{\rm s,0}},
    \label{eq:stokes}
\end{align}
where $\rho_{g,0}$ is the average density for the spherical cloud with a radius of 6686 AU, $\rho_s = 3.6~\rm g~cm^{-3}$ is the average density of the grains, and $\tau_{\rm acc} = t_{\rm acc}^{-1}$ is the timescale related to the collapse. This timescale is the most relevant to the hydrodynamic properties of the dust given that early on in their collapse the speeds in the spherical radius direction tend to be higher than their rotational speeds. As the cloud collapses and the dust grains fall towards the centre their local Stokes number decreases, since the gas density tends to increase and we do not invoke any grain growth. When they reach the disk, however, the local dynamical timescale ($\Omega$) increases because orbital rates are faster than radial speeds in a rotationally supported disk, while the gas volume density increases as well. For the majority of the grain models the Stokes number increases when the grain reaches the disk. 

The fourth simulation uses the 34$\mu$m dust grains and the same cloud initial conditions other than a larger solid body rotation rate of $\Omega_{\rm ref} = 10^{-13} {\rm s}^{-1}$, and will be known as the Ref-med model. The standard and the reference cases are known as the spread- and infall-dominated cases in \cite{Drozdovskaya2016} as they reflect the primary method for which the embedded disk grows given our (relatively high) choice of $\alpha$. The lower initial rotation rate, and subsequently lower gas angular momentum, of the standard case causes the embedded disk to start small ($\sim 10$ AU) and primarily grow through viscous spreading. While in the reference (infall-dominated) case the embedded disk starts larger ($\sim 100$ AU) and further grows through a combination of viscous spreading and subsequent envelope accretion (see Table \ref{tab:params} for final sizes). Since we assume that the dust grains begin their evolution with the same rotation rate as the gas dust falling in the standard case will tend to fall into the disk closer to the host star than grains falling in the reference case.

We summarise the important physical parameters in Table \ref{tab:params}. Included in the table are estimates of the Stokes number for each of the dust grains initially ($St_{\rm env}$) as well as at a future time when the grain has entered the embedded disk ($St_{\rm disk}$). When the grains enter the disk they encounter a higher gas density but also much faster rotation speeds which drives a higher Stokes number of each of the models.

\begin{table*}
    \centering
    Basic physical parameters\\
    \begin{tabular}{c c c c c c c}
    \hline\hline
        Run & s ($\mu$m) & $\Omega_0$ (s$^{-1}$) & $St_{\rm env}$ & $St_{\rm disk}$ & R$_{\rm gas}$ (AU) & M$_{\rm gas}$ (M$_\odot$)\\\hline
        Std-lrg &  3400 & 10$^{-14}$ & 12 & 26 & 56 & 0.05\\
        Std-med & 34 & 10$^{-14}$ & 0.12 & 1.1 & 56 & 0.05\\
        Std-sml & 0.34 & 10$^{-14}$ & 0.0012 & 0.0025 & 56 & 0.05 \\
        Ref-med & 34 & 10$^{-13}$ & 0.12 & 0.2 & 440 & 0.24\\
        \hline
    \end{tabular}
    \caption{Relevant physical parameters for the four simulations presented here. Common to all simulations is the initial cloud mass (1 M$_{\odot}$), the initial cloud sound speed (0.26 km s$^{-1}$), and $\alpha = 0.01$ in the disk model. R$_{\rm gas}$ and M$_{\rm gas}$ are the radial extent of the embedded gas disk and its mass at $t_{\rm acc}$.}
    \label{tab:params}
\end{table*}

\subsection{ Single dust trajectory }

As a first step in our analysis we present a set trajectories of the same parcel of dust in each simulation. In this way we directly compare the effects of the initial angular momentum and grain size on the overall collapse history of the dust.

\begin{figure}
    \centering
    \begin{overpic}[width=0.5\textwidth]{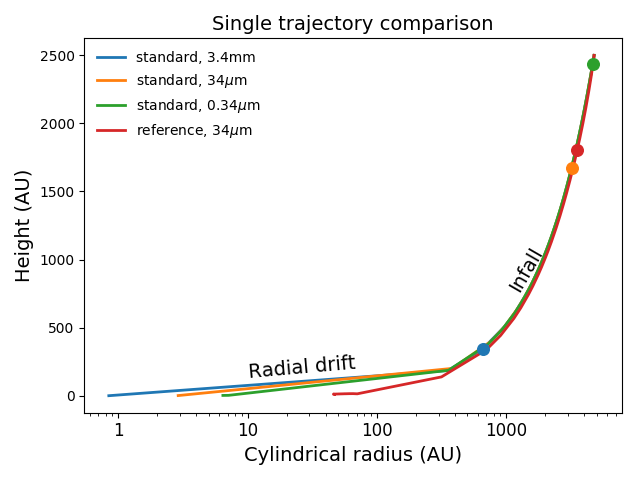}
        \thicklines
        \put(90,60){\color{black}\vector(-1,-4){5}}
        \put(75,25){\color{black}\vector(-4,-1){20}}
        \put(83,50){\rotatebox{75}{\color{black} Time}}
    \end{overpic}
    \caption{A single dust parcel trajectory starting from the same location in the envelope for all four models. Broadly speaking time flows from the top right of the figure to the bottom left. The coloured point denotes the location of the parcel in each model at a collapse time of 120 kyr. The size of the dust particles in each collapse model denotes how quickly the dust collapses, with larger grains being further along in their collapse than smaller grains. The different rotation rates of the clouds produces a small change in the envelope collapse, and grows the embedded disk more quickly - which causes a large change to the trajectory when the dust particle encounters the disk. }
    \label{fig:singletraj}
\end{figure}

In Figure \ref{fig:singletraj} we show the collapse trajectories for a parcel of dust beginning at the same location in each model. In the standard model nearly all of the different grain sizes follow a similar path, however the rate of the collapse changes as a function of grain size. This is seen in the fact that the point marking 120 kyr (nearly halfway to $t_{\rm acc}$) is farther along the collapse trajectory of the larger grains than the smaller ones. Since the larger grains are less coupled to the gas, they collapse more freely than smaller grains - which are partly kept aloft by the gas. Once the grains enter the disk their local Stokes number increases due to the faster rotating gas and higher gas densities. Radial drift becomes a dominant source of further evolution in the embedded disk, and the larger grains tend to drift to smaller radii than the smaller ones.

The higher rotation rate of the reference case keeps the collapsing dust grains farther out in the envelope as it collapses. Once in the disk the particles do not radially drift as fast as the dust in the standard model. This is partially due to the higher outward flux of the gas as the disk is being built, and partially due to the high rotational support that arises from the higher rotation rate. As a result, the general conclusion that high-rotation rate envelopes produce larger embedded disks continues to apply to the dust as much as it does to the gas - even when we consider the impact of radial drift.

\begin{figure}
    \centering
    \includegraphics[width=0.5\textwidth]{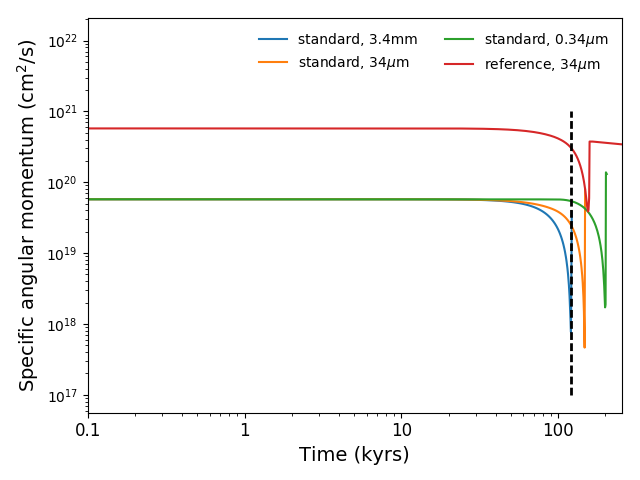}
    \caption{The evolution of the specific angular momentum - total angular momentum per mass - of the dust parcels for each of the models. The dashed line highlights a time of 120 kyr which was featured in Figure \ref{fig:singletraj}. Angular momentum is strictly conserved for the first few 10s of kyr until the dust has fallen into a higher density region of the cloud where interactions between the gas and dust begin to efficiently drain angular momentum. Once the dust reaches the disk it enters in a region with much higher angular momentum and is quickly `spun-up' by the gas. }
    \label{fig:angmom}
\end{figure}

In Figure \ref{fig:angmom} we show the evolution of the total angular momentum of the dust parcels from Figure \ref{fig:singletraj}. For the first few 10s of kyr the angular momentum is strictly conserved as the dust begins its collapse, because the relative collapse speeds of the gas and dust are initially small. After a while the dust reaches a region of the envelope with higher gas densities and has accelerated such that its speed relative to the gas is sufficiently high that it begins to more efficiently lose angular momentum via gas drag. When the dust parcels reach the disk they begin interacting with a higher density, rotationally supported gas. Hence the relative speeds between the disk gas and the dust are large, and the dust is efficiently `spun-up' in the first few time steps after it crosses the boundary of the disk. As a result the dust parcel's angular momentum suddenly spikes when it enters the disk, but then flattens out once its rotation speed is once again similar to the local gas rotation speed. While the particle is spun-up it begins to radially drift through the disk, losing angular momentum more slowly than during its collapse through the envelope. The evolution during the radial drift stage of the dust evolution can best be seen in the reference model (red line), while the dust parcels in the other models require the aforementioned extrapolation to reach a time of $t_{\rm acc}\sim 252$ kyr.

During the short timeframe where the dust regains angular momentum after passing into the disk, the gas should also lose a similar amount of specific angular momentum as the dust gains. This back reaction may, in principle, impact the flow of the gas in the embedded disk. However the total dust mass represented by a given parcel in our model is much smaller than the total mass in an annulus of gas from where the dust extracts its angular momentum - hence we do not expect there to be a significant change in the rotation rate of the gas caused by this momentum exchange.

\subsection{ Embedded gas disk formation }

\begin{figure}
    \centering
    \begin{minipage}{0.49\textwidth}
            \includegraphics[width=\textwidth]{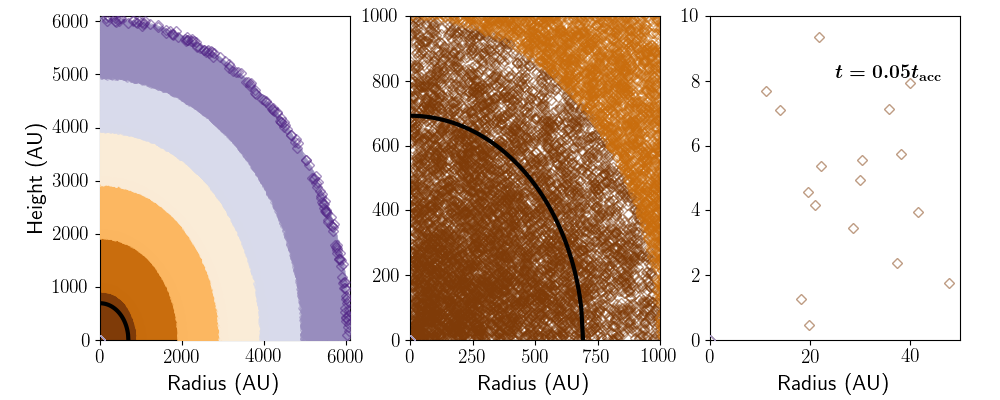}
    \end{minipage} 
    \begin{minipage}{0.49\textwidth}
            \includegraphics[width=\textwidth]{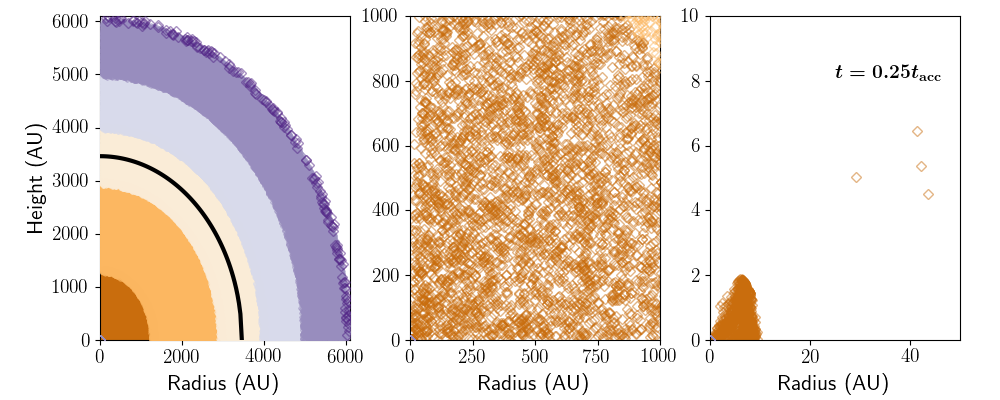}
    \end{minipage}
    \begin{minipage}{0.49\textwidth}
            \includegraphics[width=\textwidth]{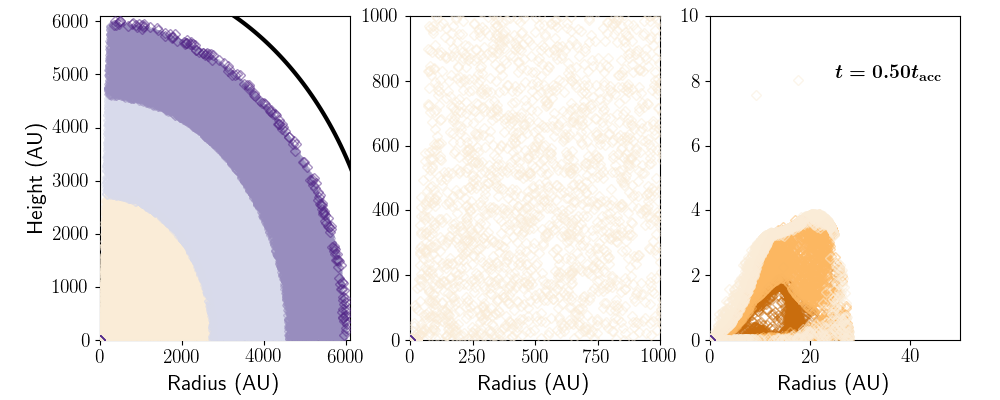}
    \end{minipage} 
    \begin{minipage}{0.49\textwidth}
            \includegraphics[width=\textwidth]{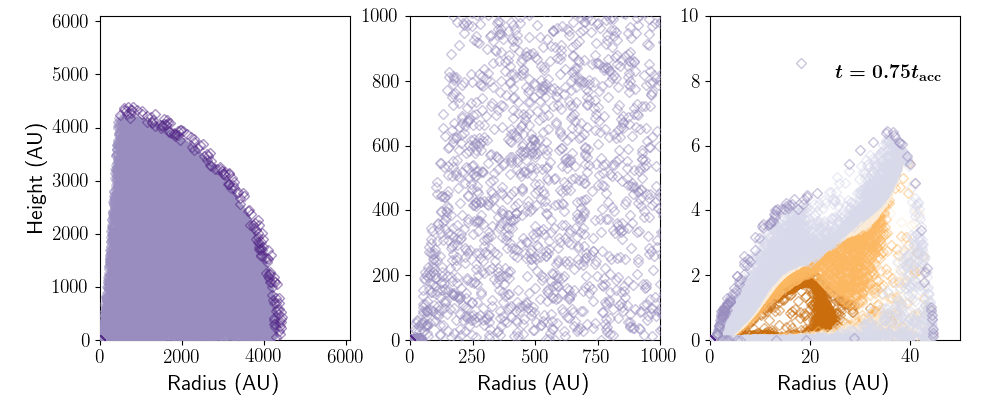}
    \end{minipage}
    \caption{ The collapse and disk build up of the gas parcels in the standard model. Each set of three panels shows the collapse at 0.05, 0.25, 0.5, and 0.75 $t_{\rm acc}$ over three length scales. The left panel of the three shows the largest, envelope, scale while the right most panel of the three shows the length scale relevant to the disk. The colour of each point denotes the initial radius range from which the parcel originates. Dark brown denotes initial radius less than 1000 AU, orange between 1000-2000 AU, etc. The embedded disk builds up from gas successively falling from higher heights as well as the viscous spreading of the disk itself. For reference, the black line shows an estimate of the location of the rarefaction wave define by $x=1$.}
    \label{fig:gas_buildup}
\end{figure}

In the following section we report the build up of the embedded dust disk however first, for reference, in Figure \ref{fig:gas_buildup} we show trajectories for the collapse of the gas particles in the standard model. The method for computing these trajectories are exactly the same as developed by \cite{Visser2009}. The colour coding of the points shows from where in the initial envelope the gas particles originated. Brown points shows initial radii less than 1000 AU, dark orange for initial radii between 1000-2000 AU, light orange from radii between 2000-3000 AU, etc. The four different sets of three panels show four different timestep in the collapse at three different length scales. The left panel of three shows the envelope length scale, the right panel of the three shows the length scale relevant to the disk, while the middle panel shows an intermediate length scale. The timesteps shown here are at $t = 0.05$, 0.25, 0.5, and 0.75 $t_{\rm acc}$.

As the gas collapses it is either lost to the host star, enters into the disk where it begins to evolve viscously, or is ejected by the outflow. As previously mentioned the embedded disk in the standard model primarily grows through viscous spreading rather than direct infall. This can be seen by the particles coming from smaller initial radii in the envelope remaining in the disk and slowly expanding their cylindrical radius. Gas parcels beginning farther out in the envelope take a longer time to reach the disk, and enter the disk at slightly higher heights and at larger radii before further expanding viscously. On the middle of each set of three panels we can see the development of the growing outflow cavity. The size and evolution of this outflow is handled very simply by parameterization \citep[see][for details]{Visser2009}. In the trajectory calculation the gas particles that enter into the outflow are immediately removed.

\begin{figure}
    \centering
    \begin{minipage}{0.49\textwidth}
            \includegraphics[width=\textwidth]{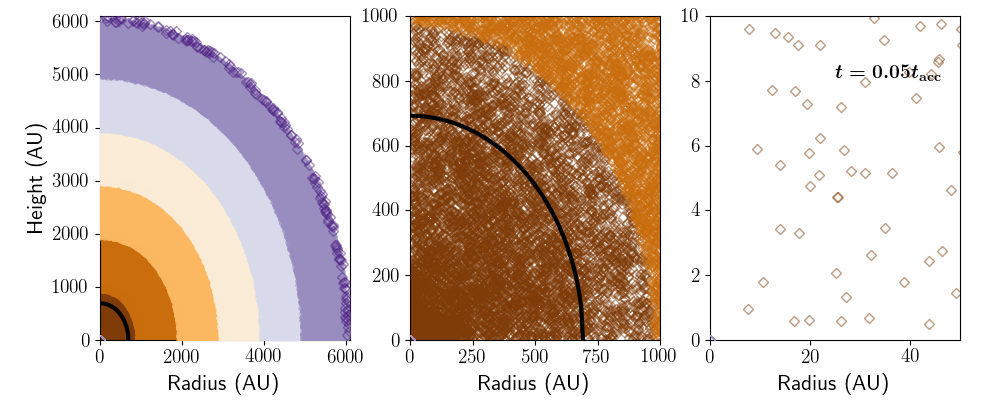}
    \end{minipage} 
    \begin{minipage}{0.49\textwidth}
            \includegraphics[width=\textwidth]{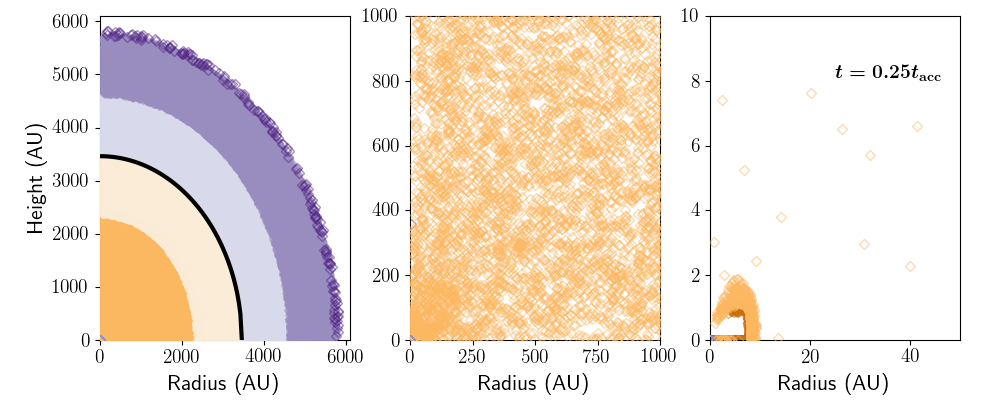}
    \end{minipage}
    \begin{minipage}{0.49\textwidth}
            \includegraphics[width=\textwidth]{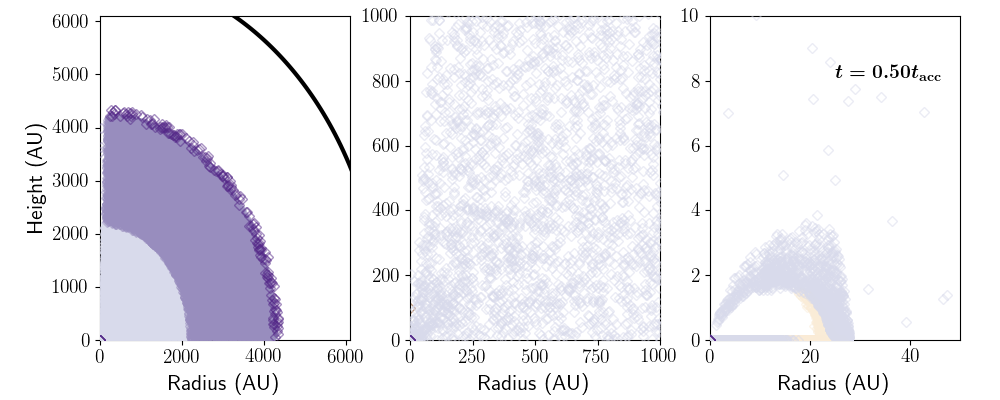}
    \end{minipage} 
    \begin{minipage}{0.49\textwidth}
            \includegraphics[width=\textwidth]{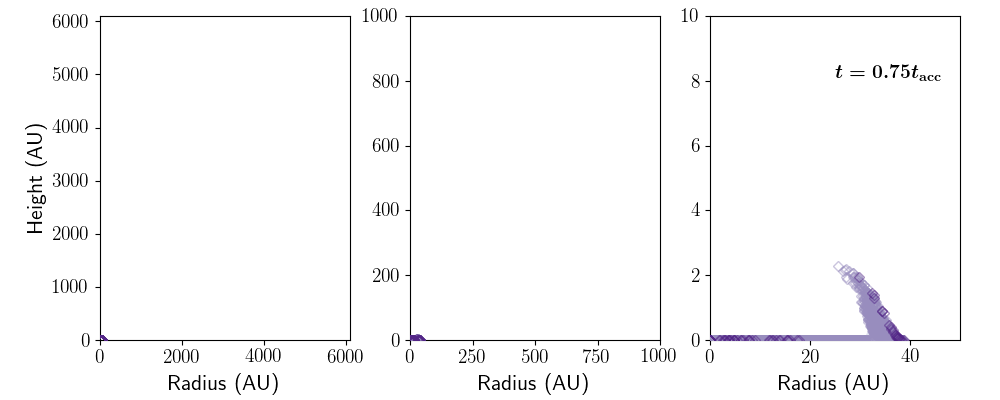}
    \end{minipage}
    \caption{The collapse of dust parcels in the Std-med model. The top left set of three panels show a very early time step in the collapse. The colour of each point is the same as in Figure \ref{fig:gas_buildup}. Each set of three panels show the same timestep at three different lengths scales. With the right-most panel in the set of three showing the length scales relevant for the embedded disk. The second, third, and forth set of three represent the simulation at 0.25, 0.5, and 0.75 $t_{\rm acc}$ respectively. Note that as the disk builds some of the dust radially drifts inward, some advect outward with the growing gas disk, and missing dust is replenished by newly infalling dust. For comparison, the black line shows an estimate of the location of the gas rarefaction wave define by $x=1$.  }
    \label{fig:smdust_disk_buildup}
\end{figure}

\begin{figure}
    \centering
    \begin{minipage}{0.49\textwidth}
            \includegraphics[width=\textwidth]{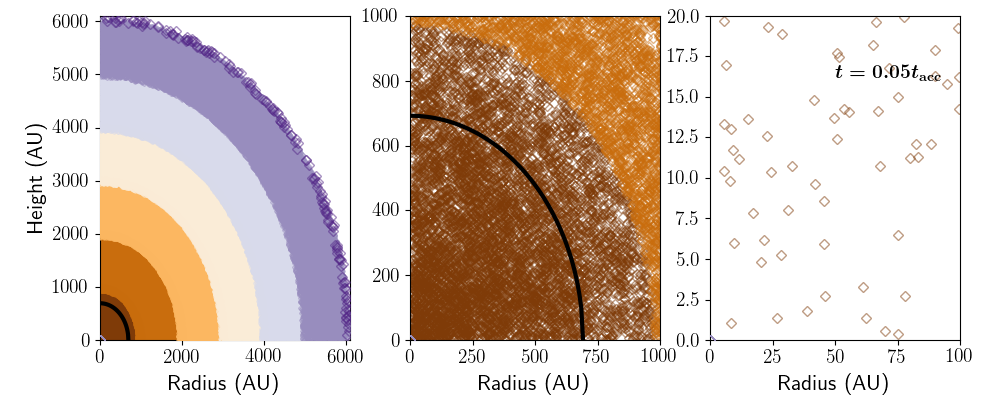}
    \end{minipage} 
    \begin{minipage}{0.49\textwidth}
            \includegraphics[width=\textwidth]{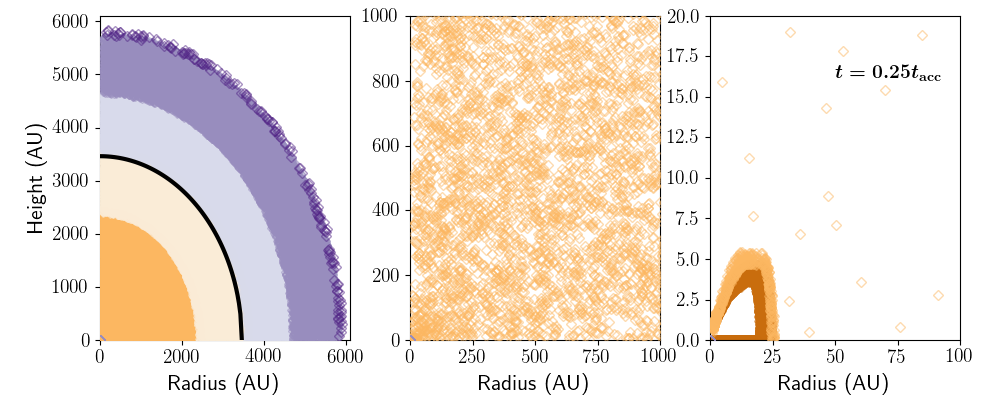}
    \end{minipage}
    \begin{minipage}{0.49\textwidth}
            \includegraphics[width=\textwidth]{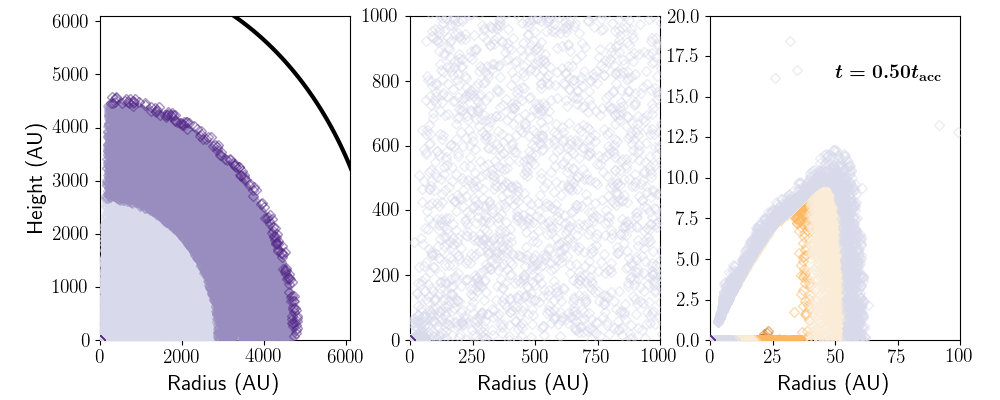}
    \end{minipage} 
    \begin{minipage}{0.49\textwidth}
            \includegraphics[width=\textwidth]{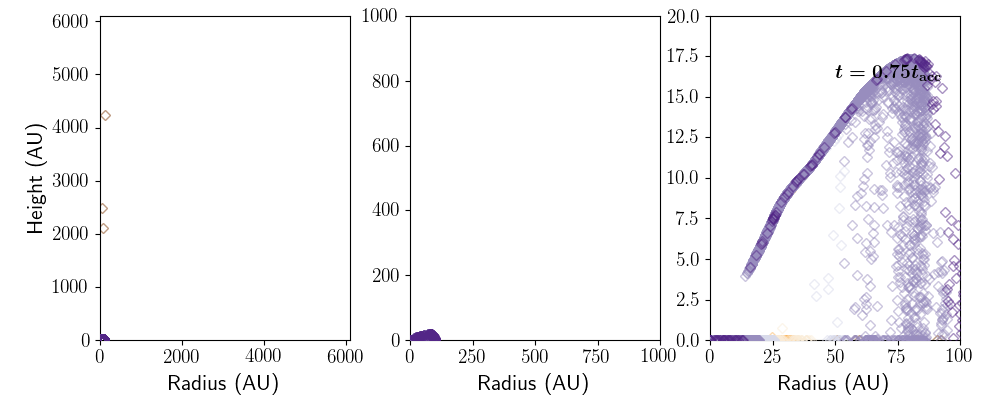}
    \end{minipage}
    \caption{ The collapse of dust parcels in the Ref-med model. The colour and timesteps are the same as in Figures \ref{fig:gas_buildup} and \ref{fig:smdust_disk_buildup}, however note the change in x-axis scale in the right most panels. While the overall collapse is over a similar timescale as in the Std-med model, the disk build up and subsequent evolution in the disk is different. The gas disk is bigger in the reference model and, on average, dust grains fall predominately at the outward edge of the disk. Viscous spreading is still present in the embedded disk, but the majority of the dust evolution is inward via radial drift. As such at the late times we find a large, extended dust ring with the majority of the dust inward of the ring along the midplane or at the upper surface of the disk. For comparison, the black line shows an estimate of the location of the gas rarefaction wave define by $x=1$. }
    \label{fig:smdust_em13}
\end{figure}

\subsection{ Dust disk formation }

\subsubsection{The standard model}

In Figure \ref{fig:smdust_disk_buildup} we show the collapse of the dust in the standard model with dust grain size of 34 $\mu$m (called the Std-med model in table \ref{tab:params}) at $t=$ 0.05, 0.25, 0.5, and 0.75 $t_{\rm acc}$. The colour of the points represents the same colours used in Figure \ref{fig:gas_buildup} to show how the rate of collapse varies between the gas and dust.  

By comparing Figures \ref{fig:smdust_disk_buildup} to \ref{fig:gas_buildup} we confirm that the dust falls faster than the gas due to its lack of pressure support. This can be seen in the second set of panels ($t=0.25 t_{\rm acc}$) in Figures \ref{fig:gas_buildup} and \ref{fig:smdust_disk_buildup} where the small gas disk is made up of gas parcel originating from between 1000-2000 AU (dark orange) while the dust component of the disk originates from between 2000-3000 AU (light orange). In addition, note the relative location the coloured parcels with respect to the gas rarefaction wave (solid black line) in the left-most panel of the second set. 

At later times, in the third set of panels ($t=0.5 t_{\rm acc}$), the dust disk is made primarily of dust parcels originating from between 3000-5000 AU (off-white and light blue points) while the gas disk still contains gas parcels originating from between 1000-2000 AU. This reflects the impact of radial drift which drives the inward evolution of dust particles in the embedded disk. This ensures that the bulk of the embedded disk is made up of fresh dust grains falling from the envelope, since the colour of the embedded disk changes with every timestep. At intermediate timesteps (0.25, 0.5 $t_{\rm acc}$; second and third sets of three panels, respectively) the dust disk follows a similar extent as the gas disk, but by later times (0.75 $t_{\rm acc}$; bottom set of three panels), due to radial drift and settling, the dust parcels are not as extended in height as are the gas parcels. Unlike in the gas simulation, there are few dust parcels remaining in the envelope by 0.75 $t_{\rm acc}$, and so at late times the mass flux onto the embedded disk is dramatically reduced.

Some similar features are seen in the collapse calculations for the standard model with 3.4 mm grains (Std-lrg) and with the 0.34 $\mu$m grains (Std-sml) - see Figures \ref{fig:dust_disk_buildup} and \ref{fig:vsmdust_disk_buildup} respectively. The dust grains in the Std-lrg model evolve very quickly and only develop a very small, compact dust disk before that radially drifts into the host star. The dust in the Std-sml model, on the other hand, follows the gas more closely than either of the Std-med and Std-lrg grain models, due to its higher coupling. As a result there is still a significant amount of dust in the envelope at late times which will help to maintain dust mass in the Std-sml model. Its distribution also more closely follows the extent of the gas disk. In the Std-sml model we see some parcels being captured into the outflow cavity. In our analysis we generally ignore dust parcels that enter into the outflow assuming these are carried outward and not inward.

\subsubsection{The reference model}

In Figure \ref{fig:smdust_em13} the reference model (Ref-med) is shown over the same timesteps as Figure \ref{fig:smdust_disk_buildup}. Since this model uses the same dust grain sizes as the Std-med model the collapse occurs over a similar timescale. The disk build-up, however, is different mainly because the size of the gas disk in the reference model is larger than in the standard model. Because of the size of the disk, we find that settling and radial drift are both slightly less efficient for the 34 $\mu$m in the reference model, as can be seen by comparing the colour of the dust parcels contributing to the embedded disk at intermediate times. At both 0.25 and 0.5 $t_{\rm acc}$, there are dust parcels that originated between 1000-2000 AU and 2000-3000 AU respectively in the Ref-med embedded disk, while dust parcels originating from these radii have completely disappeared in the Std-med model. At 0.75 $t_{\rm acc}$ the envelope has been depleted of dust, however the embedded disk remains large, with many dust grains still in the process of settling towards the midplane. 

Overall we see that there may be a sweet spot for planetesimal formation with particular dust grain sizes given their collapse trajectories. Large grains fall very quickly, then settle and radially drift into the host star so that they may not build up enough dust density to spur the streaming instability. Small grains that are highly coupled to the gas stay largely extended, and are much slower to accreted on to the embedded disk, they also do not settle to the midplane as efficiently. Medium sized grains act as an intermediary of the two extremes, collapsing quicker than the small grains because of their slight decoupling with the gas, and settling efficiently enough to possibly grow a density enhancement along the midplane. Furthermore the radial drift rate is slower than the large grains so that the density enhancements last longer in the embedded disk.

\subsection{ Total mass evolution }

One of the important observables in protostellar systems is the total dust mass contained with the embedded disk. In this section we report the evolution of the total embedded dust and gas mass in our models, and compare the embedded dust mass to the observational results of \cite{Tychoniec2020}. The total embedded dust mass is computed by summing the assigned mass (as explained in section \ref{sec:method_assignment}) of all particle that exist within the embedded disk surface at a given time. The embedded gas mass is computed by integrating the analytic surface density profile that describes the disk out to its outer radius.

\ignore{
These are achieved \citep[for instance in ][]{Tychoniec2020} by removing extended (sub)-millimetre flux which is assumed to originate from dust leftover in the envelope. With the removal of the extended flux, the remaining flux is converted to a dust mass using the typical relation \citep{Hildebrand1983}:\begin{align}
    M_{\rm dust} = \frac{D^2F_\nu}{\kappa_\nu B_\nu(T_{\rm dust})},
\end{align}
where $T_{\rm dust}$ is assumed to be 30 K for protostellar systems. We compare the mass evolution of our collapse models to the observational results of \cite{Tychoniec2020}.
}

\begin{figure}
    \centering
    \includegraphics[width=0.5\textwidth]{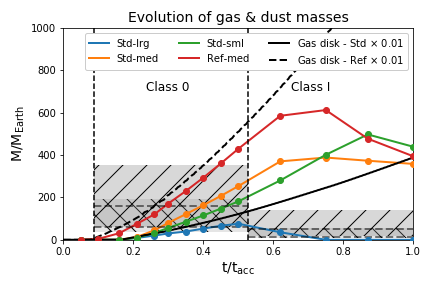}
    \caption{The time evolution of the total dust mass inside of the embedded disk in each model over one $t_{\rm acc}$ ($\sim$252 kyr). The grey bands show the median disk masses for ALMA (right hatch) and VLA (left hatch) detected protostars reported by \cite{Tychoniec2020} along with the confidence interval of the median. The ALMA-inferred masses are likely lower limits due to the dust being optically thick, while the VLA-inferred masses are more likely to be optically thin and use realistic opacities assuming the existence of large grains. The vertical dashed lines show the time where the embedded disk first appears (classifying the system as Class 0) and when the disk gas mass exceeds the remaining gas mass in the envelope (classifying the system as Class I). Both Class 0 and Class I objects were included in \cite{Tychoniec2020} and we have separated them to compute individual medians and confidence intervals. For comparison, we include the mass evolution of the disk gas in both the standard (black solid) and reference (black dashed) models - scaled by the ISM dust-to-gas ratio. } 
    \label{fig:mass_evo}
\end{figure}

Figure \ref{fig:mass_evo} presents the evolution of the gas and dust mass within the embedded disk. The black curves denote the temporal evolution of the embedded gas disk mass for the standard (solid line) and reference (dashed line) models. These masses are in units of Earth masses and scaled by the ISM dust-to-gas ratio to better compare the curves to the dust masses (coloured lines). Since the embedded disk in the reference model is physically larger (recall Table \ref{tab:params}), its mass also grows larger than the embedded disk in the standard model. 

We define the dust mass by keeping track of all dust parcels that are found inward of the disk surface at a given time, and summing all of their dust masses together. Since the inner disk radius does not extend all the way to $R=0$, dust that radially drifts past the inner boundary of the disk is considered lost to the host star. We find that the embedded dusty disk mass slowly climbs over nearly the whole lifetime for the models with grain sizes $\leq 34~\mu$m. The largest grains grow to a maximum mass of nearly 100 M$_\oplus$ at $t\sim 0.5 t_{\rm acc}$ before rapidly draining due to radial drift. The smaller grains only begin to have a negative flux of dust mass at late times when their envelope has largely dissipated. The left-most vertical dashed line shows the location where the embedded disk first appears in the simulation, we denote this time as the emergence of our Class 0 object.    

The Class I phase of our simulated systems is defined as the point where the gas mass in the embedded disk is equal to the remaining gas mass in the envelope. This occurs just over 0.5 $t_{\rm acc}$ and is denoted by the right vertical dashed line. During this phase, the Std-med and Std-sml models see sustained dust mass flux from their envelopes because grains of this size are more coupled to the gas than in the Std-lrg model which delays their collapse with respect to the larger grain sizes. Since the 34 $\mu$m grains are less coupled than the 0.34 $\mu$m grains, the Std-med model stalls its growth earlier than the Std-sml model. The grains in the reference model show a faster built up than embedded disks in the standard models. Primarily because the gas disk grows larger and faster in the reference model, more dust grains are caught early in the collapse of the Ref-med model leading to the heavier embedded disk. Furthermore, since the settling and radial drift are less efficient in the Ref-med model than in the standard models, the disk holds onto its dust longer, allowing for a larger build up of mass.

Comparing the evolution of the dust mass to the gas mass in each of the model we can already see that the reference model may struggle to make planetesimals, since its global dust mass is always less than the mass of the embedded disk gas - scaled by the ISM dust-to-gas ratio. The standard models on the other hand, appear to favour the generation of planetesimals because their dust masses in the Std-med and Std-sml models are always larger than the scaled embedded disk gas mass. We will investigate this further in the following sections, since the streaming instability tends to be a \textit{local} process the local dust-to-gas ratio is much more relevant to its physics than a global one.

Included in Figure \ref{fig:mass_evo} is the median dust mass (grey horizontal dashed line) and its confidence interval (hashes) inferred by \cite{Tychoniec2020} for a population of protostellar systems studied with ALMA and the VLA. The difference between their inferred median mass and confidence intervals comes from the different dust opacities used in converting luminosity to mass. We discuss this in more detail in Section \ref{sec:interpmass}.  We use the python package \textit{lifelines}\footnote{https://lifelines.readthedocs.io} to compute the median dust masses and their confidence interval. This population includes both Class 0's and Class I's which we separate in order to compute the median dust mass for each protostellar type. The observational data brings to light a particular feature that our simulations do not tend to reproduce - that Class 0 objects tend to have higher dust masses than their older Class I cousins. The discuss our interpretations of this in section \ref{sec:discuss}.

\subsection{ Dust-to-gas ratio evolution }

\begin{figure*}
    \centering
    \begin{minipage}{0.49\textwidth}
            \includegraphics[width=\textwidth]{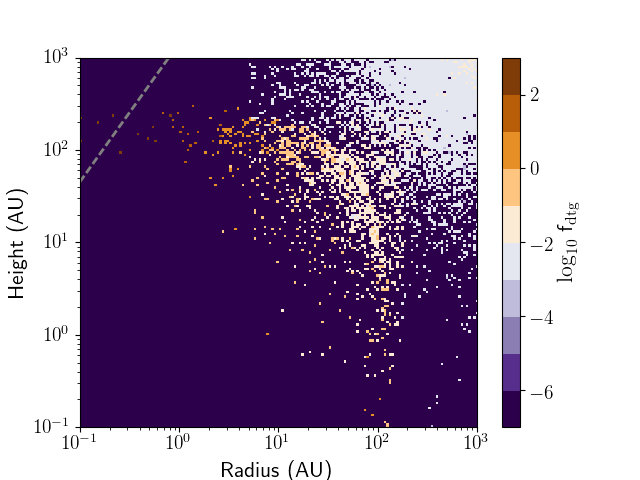}
    \end{minipage} 
    \begin{minipage}{0.49\textwidth}
            \includegraphics[width=\textwidth]{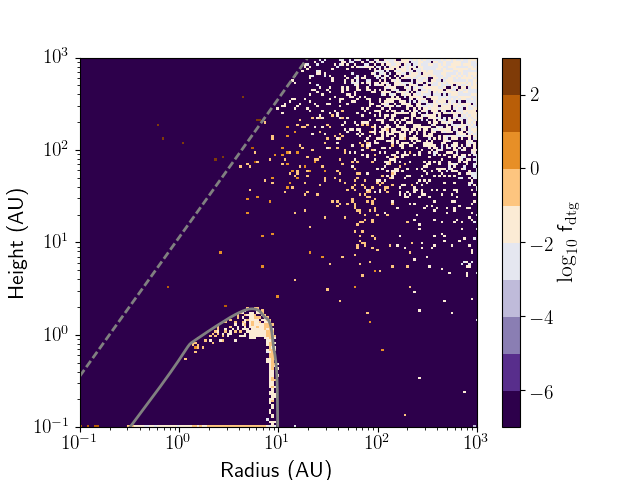}
    \end{minipage}
    \begin{minipage}{0.49\textwidth}
            \includegraphics[width=\textwidth]{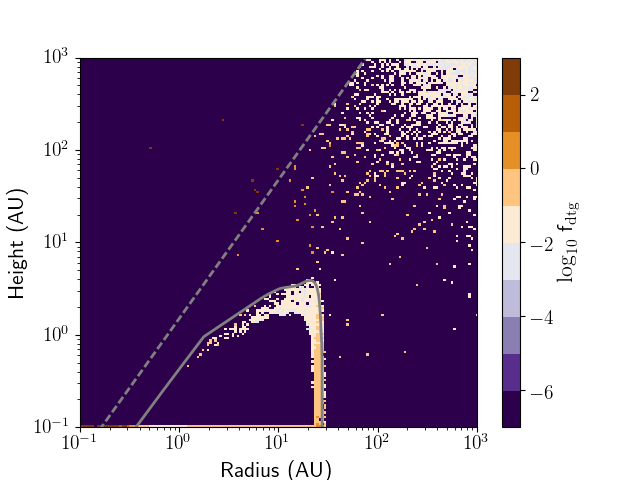}
    \end{minipage} 
    \begin{minipage}{0.49\textwidth}
            \begin{overpic}[width=\textwidth]{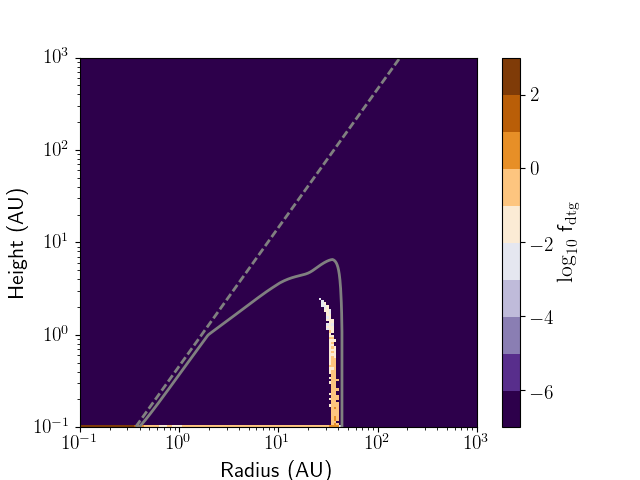}
            \thicklines
            \put(25,25){\color{gray}\vector(1,-1){17}}
            \end{overpic}
    \end{minipage}
    \caption{ The dust-to-gas mass ratio in the Std-med collapse model. The colour scale shows log of the local dust-to-gas ratio, hence blue denote dust-to-gas ratios below the ISM while orange denotes dust-to-gas ratios above the ISM. The dashed lined shows the extend of the outflow in the gas model while the solid line shows the embedded disk's surface. Note that we assume that planetesimal formation occurs largely along the midplane - highlighted by the grey arrow. We show the midplane dust-to-gas ratio below.}
    \label{fig:smdust_gas-to-dust}
\vspace{5pt}
    \begin{minipage}{\textwidth}
    \begin{overpic}[width=0.9\textwidth]{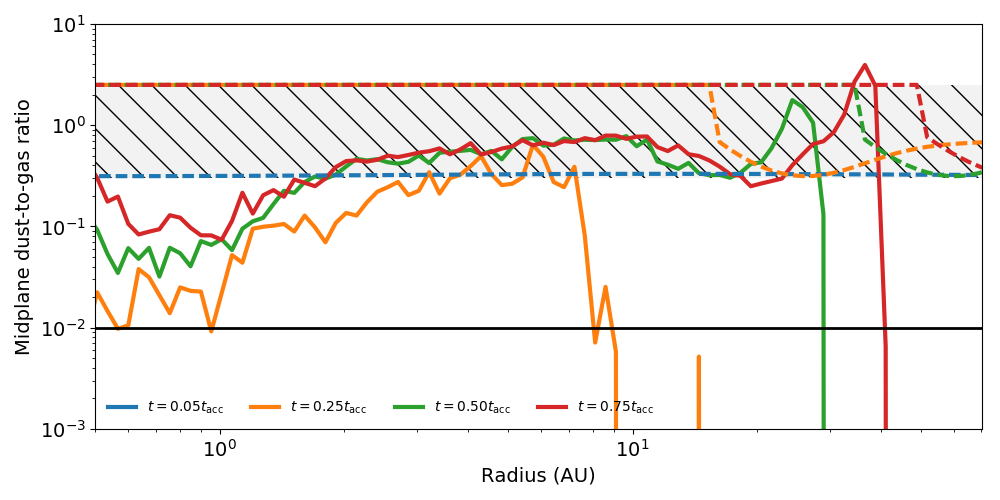}
        \put(50,18){\color{black}\Large ISM}
        \put(15,39){\color{black}\Large Threshold for streaming instability}
    \end{overpic}
    \caption{The midplane dust-to-gas ratios for the Std-med collapse model at the same time steps as shown in Figure \ref{fig:smdust_gas-to-dust}. At the earliest times (solid blue curve) the dust disk has yet to form and does not appear on this plot. The black line shows the ISM dust-to-gas ratio which represents the initial conditions in the envelope, while the coloured dashed lines show the streaming instability threshold at each timestep. The hashed region is an estimate of the streaming instability threshold based on \cite{YG05}, while the dashed lines show the threshold based on \cite{LY2021}. Only in the last timestep is the threshold met along the midplane of the embedded disk.}
    \label{fig:midplane_smdust}
    \end{minipage}
\end{figure*}

\begin{figure*}
    \centering
    \begin{minipage}{0.49\textwidth}
            \includegraphics[width=\textwidth]{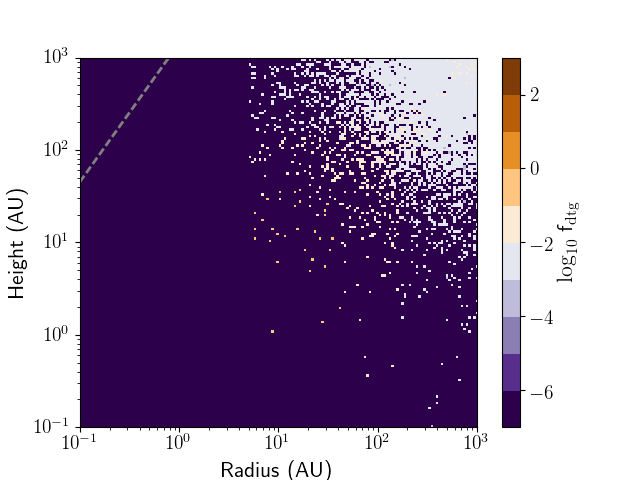}
    \end{minipage} 
    \begin{minipage}{0.49\textwidth}
            \includegraphics[width=\textwidth]{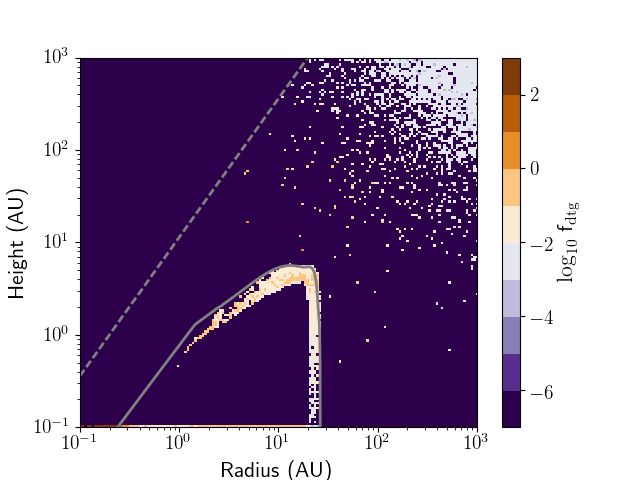}
    \end{minipage}
    \begin{minipage}{0.49\textwidth}
            \includegraphics[width=\textwidth]{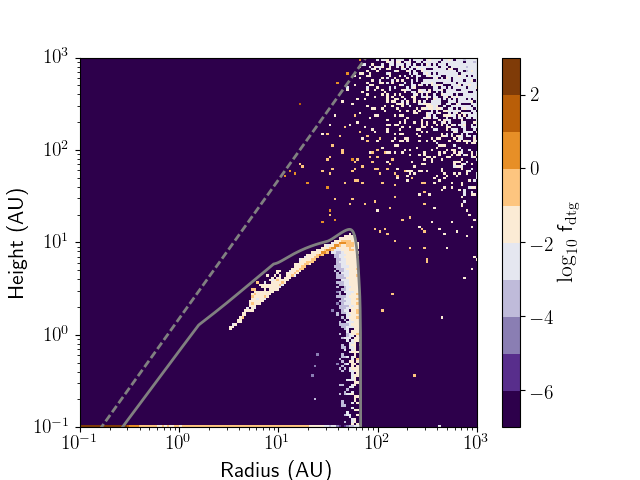}
    \end{minipage} 
        \begin{minipage}{0.49\textwidth}
            \begin{overpic}[width=\textwidth]{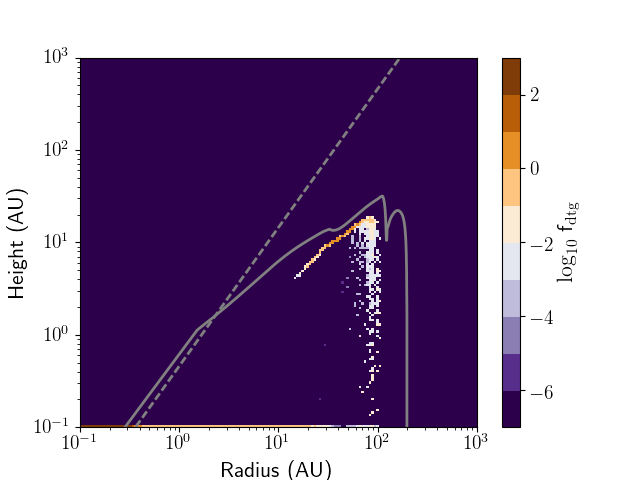}
            \thicklines
            \put(25,25){\color{gray}\vector(1,-1){17}}
            \end{overpic}
    \end{minipage}
    \caption{ Same as Figure \ref{fig:smdust_gas-to-dust} but for the reference model Ref-med. Here we see that settling is less efficient than in the standard model with the majority of the dust grains remaining near the upper surface of the disk. There are regions just outward of the disk surface along the midplane that show sufficiently high dust-to-gas ratios. These regions, however, do not survive to the last snapshot shown here. Planetesimal formation predominantly occurs along the midplane, highlighted by the grey arrow in the forth panel, and shown below.}
    \label{fig:smdust_em13_gas-to-dust}
\vspace{5pt}
    \begin{minipage}{\textwidth}
    \begin{overpic}[width=0.9\textwidth]{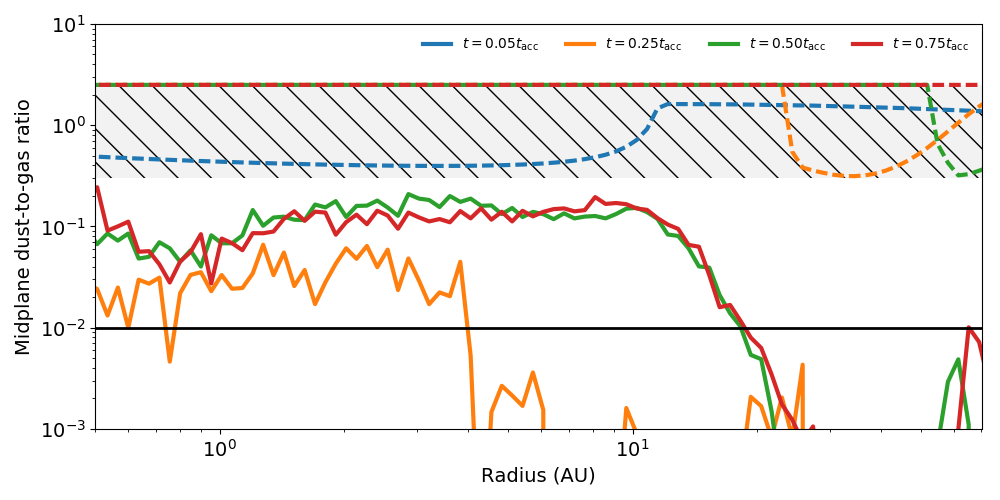}
        \put(50,18){\color{black}\Large ISM}
        \put(15,39){\color{black}\Large Threshold for streaming instability}
    \end{overpic}
    \caption{The midplane dust-to-gas ratios for the Ref-med collapse model at the same time steps as shown in Figure \ref{fig:smdust_em13_gas-to-dust}. The lines and hashed regions have the same definition as in \ref{fig:midplane_smdust}. Here we see that the dust-to-gas ratio never exceeds the threshold for the onset of the streaming instability at any of the shown timesteps.}
    \label{fig:midplane_em13dust}
    \end{minipage}
\end{figure*}

In Figure \ref{fig:smdust_gas-to-dust} we show the evolution of the dust-to-gas ratio during the collapse of the Std-med model at the same time steps as in Figure \ref{fig:smdust_disk_buildup}. The gas density is computed using the smooth solution of the \cite{Visser2009} model (rather than using the trajectories seen in \ref{fig:gas_buildup}) and the dust density is computing using the collapse trajectories shown in Figure \ref{fig:smdust_disk_buildup} and as discussed in section \ref{sec:method_assignment}. The colours in the figure denote the local dust-to-gas ratio in the simulation. More orange points show progressively more dust-rich regions of the system while more blue points show more gas-rich regions.

The set of four panels shows the dust-to-gas ratio in the system at $t=$ 0.05, 0.25, 0.5, and 0.75 $t_{\rm acc}$ in the top-left, top-right, bottom-left, and bottom-right positions of the Figure respectively. The grey dashed line shows the edge of the gas outflow, while the solid grey line shows the surface of the embedded disk which is defined as the location where the gas pressure in the disk model exceeds the gas pressure in the cloud. When the gas disk forms it immediately acts as a trap for the falling dust parcels, with the majority of the dust being trapped near the outer and upper edge of the disk surface. 

The dust that is in the apparent traffic jam at the upper edge of the embedded disk is only temporarily in that region. The speeds of the dust parcels in this region are sub-sonic, with typical speeds up to a few 100s of cm s$^{-1}$ (compared to the typical sounds speed there of $\sim 10^4$ cm s$^{-1}$). So while the inward velocity of the collapsing dust parcels is greatly slowed as they transition from the envelope to the disk, they return to accelerating towards the midplane after a short residence at the top of the disk atmosphere. This short residence time could provide enough time for grain growth to occur in the top of the disk atmosphere, which we investigate in the discussion section below.

Since we ignore the impact of turbulent diffusion in computing the dust density in the embedded disk, there are regions where no dust mass can be found. As such the three major dust components in our models are the aforementioned dust trap at the disk surface, the dust along the midplane, and the still falling dust in the envelope. In reality we would expect that turbulent diffusion would mix the dust as it passes from the disk surface to midplane, and would tend to suppress the settling of the smaller grains. We further discuss the role of turbulence on our model in a later section.

\subsubsection{Midplane dust-to-gas ratio}

The relevant portion of the model where we assume planetesimal formation occurs is near the midplane of the embedded disk. This region is small in Figure \ref{fig:smdust_gas-to-dust} (shown by an arrow in the forth panel), and hence we highlight this region with Figure \ref{fig:midplane_smdust}. The colour of the solid and dashed lines denotes the particular timestep in the simulation. The solid lines show the dust-to-gas ratio along the midplane of the embedded disk, while the dashed lines show the threshold dust-to-gas ratio needed for the streaming instability as computed in equation \ref{eq:fit}. At 0.05 $t_{\rm acc}$ there is insufficient dust along the midplane to be shown in the Figure. The general structure of the dust-to-gas ratio along the midplane of the disk includes a strong peak in dust density near the outer edge of the disk, driven by the recent accretion of dust, a relatively high density of dust between 1 and 10 AU driven by the flux of dust from the outer disk through radial drift, and settling from the upper edge. Finally inward of 1 AU the dust density slowly builds with time as it is fed by inward drifting dust grains. The gas density in the inner most region along the midplane is slowly vacated by the growing inner cavity and thus radial drift becomes inefficient there.

By 0.25 $t_{\rm acc}$ there is already enough dust along the midplane of the disk to raise the dust-to-gas ratio above the ISM value by an order of magnitude outside of 1 AU. By 0.5 $t_{\rm acc}$ the outer 10 AU of the embedded disk (20-30 AU) reaches a dust-to-gas ratio of unity, however it does not exceeds the threshold needed for the streaming instability at that time. By 0.75 $t_{\rm acc}$ the dust-to-gas ratio exceeds the threshold needed for the streaming instability in a region between $\sim$30-40 AU. 

We reiterate that generally speaking the dust at a given timestep at the top and outer edges of the disk has only recently been accreted. This can be seen in Figure \ref{fig:smdust_disk_buildup} because there is little mixing between dust grains originating from different radii. Dust parcels originating from smaller radii have radially drifted inward before parcels from larger radii arrive. This ensures that the high dust-to-gas ratio regions at intermediate to late times are caused by the arrival of new dust parcels, rather than the outward diffusion of dust parcels that had previously entered the disk. This distinction is important, because once dust parcels are in a region of high dust-to-gas ratio they presumably would begin to produce planetesimals, thus making them unavailable for further evolution. We do not currently take this sink into account, since our interest here is to show that high dust-to-gas regions in the disk are possible given drift-driven deviations in collapse trajectories between gas and dust.

\subsubsection{Grain size dependence}

The dependence on the size of the dust grain is as expected, and the results of the Std-lrg and Std-sml models are presented in Figures \ref{fig:dust_gas-to-dust} and \ref{fig:vsmdust_gas-to-dust} respectively. The large grains evolve quickly and build up a dust-to-gas ratio no larger than a few factors of 10$\times$ the size of the ISM dust-to-gas ratio. Further more, from Figure \ref{fig:midplane_dust} we see that while the dust disk initially grows from $t = 0.25 t_{\rm acc}$ to $t = 0.50 t_{\rm acc}$, by $0.75 t_{\rm acc}$ the dust disk has radially drifted inward of about 0.8 AU - effectively disappearing. 

The very small grains on the other hand are more coupled to the gas and hence show much slower evolution than either the Std-lrg or Std-med models. As such the midplane dust-to-gas ratio takes longer to develop, as seen in Figure \ref{fig:midplane_vsmdust}, the radial drift has not reached an equilibrium with the accretion onto the disk which causes the dust density between 1 and 10 AU to constantly grow over the collapse of the system. The inner disk ($< 1$ AU) also takes longer to collect dust. The slower radial drift speed causes the outer disk to become very enhanced in the dust - resulting in a dust-to-gas ratio of almost 10. Such a large mass ratio likely breaks our assumption that the dust does not have any back reaction on the evolution of the gas, however for our simple experiment we ignore the possibility of these effects.

\subsubsection{The reference model}

In Figure \ref{fig:smdust_em13_gas-to-dust} we show the evolution of the dust-to-gas ratio for the Ref-med model. As already pointed out, this model has less efficient dust settling which results in lower dust-to-gas ratios along the midplane. However since the disk in the Ref-med model is large, more dust is captured by the growing embedded disk than in the Std-med model. The traffic jam near the disk edge, however, occurs at a lower density in the Ref-med model and thus there is less dust being captured there. Looking at the dust-to-gas ratio along the midplane, see Figure \ref{fig:midplane_em13dust}, the radial evolution of the dust is clearly different in this model than in the standard models. Unlike in the standard models, the dust density along the disk midplane grows nearly uniformly over all radii. Furthermore, while the gas disk in the reference model is much larger than in the standard model, the majority of the dust mass along the midplane extends to roughly the same radii (30-40 AU) as it does in the standard model. Hence it appears that while the amount of angular momentum initially in the cloud has a strong impact on the final size of the gas disk, the final size of the dust disk seems to be less sensitive on $\Omega_0$.

\section{Discussion} \label{sec:discuss}
\subsection{ Available mass for planetesimal formation }

\begin{figure}
    \centering
    \includegraphics[width=0.5\textwidth]{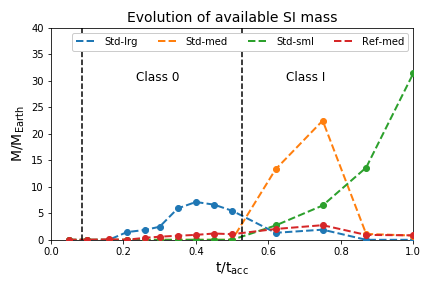}
    \caption{The temporal evolution of the mass available for the streaming instability. The streaming instability is most efficient for the 34 $\mu$m grains as the Std-med model transitions into the Class I phase of its evolution. }
    \label{fig:si_available_mass}
\end{figure}

Our main focus is to analyse whether the differential collapse between the dust and gas can lead to dust enhancements in the embedded disk that can lead the generation of planetesimals. To find what quantity of dust mass is available for the production of planetesimals we compute the threshold dust-to-gas ratio for each of the grid cells that are near the midplane of our embedded disks (within $z < 0.2$ AU). We compute the Stokes number in the gas model for grains of each size and the corresponding threshold dust-to-gas ratio in every grid cell near the midplane in each model. Next we find every grid cell where the dust-to-gas ratio exceeds the threshold ratio in equation \ref{eq:fit}, for $z < 0.2$ AU, and find all dust parcels that exist in the corresponding grid cells. Finally we compute the total mass of all of the dust parcels contributing to the grid cell's dust density to estimate the amount of mass that might be transformed to planetesimals due to the streaming instability.

Figure \ref{fig:si_available_mass} presents the total amount of dust mass that is in regions of the disk where the clumping threshold ratio is reached. The Ref-med model shows the least efficient clumping, with the least amount of dust mass available for the streaming instability over nearly the whole simulation. While the gas disk is the largest in the reference model, it shows inefficient dust settling compared to the standard models, keeping most of the dust mass off of the midplane. As a result, the dust-to-gas ratio does not get sufficiently high throughout the disk in order drive the streaming instability. 

The standard models show different available dust masses, dependent on their dust size. During the Class 0 phase, only the Std-lrg model produces regions in its embedded disk where planetesimals can be formed. From the figure it is clear that this reservoir would either be quickly used up to form planetesimals, or lost to radial drift - which is what occurs in our simulation. The large grains can produce a maximum of about 7 M$_\oplus$ of planetesimals in the Class 0 phase before the dust mass is exhausted. Meanwhile the Std-sml and Std-med models can both only start generating planetesimals during the Class I phase, because of the amount of time needed for dust to build up in the embedded disk. During the Class I phase, the Std-med model can produce a maximum of 22 M$_\oplus$ before it too loses its dust reservoir to radial drift. Only during the last $\sim50$ kyr does the dust in the Std-sml model begin to produce planetesimals, and could proceed to do so as the system transitions from Class I to II (at $t > t_{\rm acc}$). This model has a maximum of 35 M$_\oplus$ available for the generation of planetesimals near the end of the Class I phase.

The amount of planetesimals produced through the streaming instability - between 7 and 35 M$_\oplus$ is confined to a relatively small ribbon in the embedded disk (recall Figure \ref{fig:midplane_smdust}). As such it is possible that this population of planetesimal could lead to the first collapse of a large proto-planetary embryo. This mass range should be viewed as the minimally produced planetesimal mass in embedded disks. On the high mass end of the above range the proto-planetary embryo is on the order of the pebble isolation mass and would thus produce a pressure bump in the embedded disk which could contribute to dust trapping and further planetesimal formation \citep[as explored in][for example]{Lenz2019,Eriksson2020}. On the low-mass end, the embryo would be sufficiently massive to efficiently accrete pebbles via pebble accretion \citep[for ex. as discussed in][]{Bitsch2015} which would increase its mass. This growing planet would eventually reach the pebble isolation mass and similarly cause a pressure bump and dust trapping in the embedded disk. So while this initial generation of planetesimal mass is small with respect to the total mass in the embedded disk (recall Figure \ref{fig:mass_evo}), it could lead to further efficient planetesimal formation and the production of a system of planets.

\subsection{ Interpreting the dust mass evolution of young systems }\label{sec:interp}

\begin{figure}
    \centering
    \includegraphics[width=0.5\textwidth]{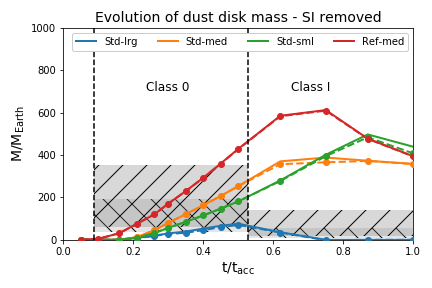}
    \caption{The evolution of the dust mass, while removing the mass available for the streaming instability. In doing so, we can more closely match the Class I disk masses found in the observations (grey regions, same as in Figure \ref{fig:mass_evo}. The Ref-med model finds almost no generation of planetesimals due to it having very little dust mass along the disk midplane (recall Figure \ref{fig:smdust_em13_gas-to-dust}).}
    \label{fig:interp_mass_evo}
\end{figure}

Figure \ref{fig:interp_mass_evo} reproduces Figure \ref{fig:mass_evo} (with solid coloured curves), and removes the mass that was available to the streaming instability which results in the dashed coloured curves - replicating the production of planetesimals from the available dust mass. Here we assume that the streaming instability is efficient enough to use all of the available mass within a few orbits, so that at each timestep we remove from the total dust mass computed in the left panel at any given timestep. Assuming that dust from previous time steps is no longer available in the embedded disk is partly supported by our previous observation that the dust parcels in the disk at a given time are largely accreted recently from the envelope. As such we assume that dust in the embedded disk is removed either by radial drift or by planetesimal formation. From figure \ref{fig:interp_mass_evo} it is clear that the production of planetesimals has only a small impact on the total dust in the embedded disk in our model.

The streaming instability-subtracted final (at $t = t_{\rm acc}$) masses range between $\sim 300-400$ M$_\oplus$ for the Std-med, Std-sml, and Ref-med models. This mass range represents the top $\sim 20\%$ of the observed Class I disks in \cite{Tychoniec2020}, which could imply that our model represents systems that are on the higher end of mass in the observed population. However our initial envelope mass of 1 M$_\odot$ (with an additional 1\% of dust mass) is near to the peak in the typical Clump Mass Function \citep[CMP,][]{Motte1998,Testi1998,K2010}, which likely represents the distribution of initial masses of star-forming clumps. Hence it is more likely that our initial setup represents a typical cloud. 

With that said, our final stellar mass is roughly 0.7 M$_\odot$, representing a star formation efficiency of about 70\%. This is much higher than the typical 30\% that is seen in comparing the CMP to the Initial Mass Function. This suggests that we are missing some inefficiency in the collapse phase of star-formation which is likely related to our simple treatment of the outflow. Whether dust is entrained in the outflow remains an open question, however given the missing inefficiencies in the evolution of the gas it is likely that our final embedded dust disk masses are also on the heavy side.

\subsubsection{Are Class 0 or Class I embedded disks more massive?}\label{sec:interpmass}

One of the primary unknowns in interpreting continuum emission data as it pertains to the amount of dust mass in young stellar systems is the opacity law that governs the radiative transfer in the cloud. \cite{Tychoniec2020} uses the opacity for their ALMA emission (at 1.3 mm) of \cite{Ansdell2016} in order to directly compare to available Class II dust mass data. Their value of 2.3 cm$^2$g$^{-1}$ is on the higher end of possible dust opacities computed by \cite{Draine2006}. Further more, \cite{Ribas2020} showed, through Spectral Energy Distribution modelling, that the dust masses observed by ALMA could be as high as 3$\times$ the masses inferred by \cite{Tychoniec2020}. In addition, \cite{Zhu2019} have argued that dust scattering, as an additional source of opacity, will generally cause underestimations of continuum emission dust masses seen by ALMA. The computed masses off \cite{Tychoniec2020} from ALMA continuum should thus be considered the minimum dust masses of young systems in their survey, which could improve the comparison between our models and observational data.

The dust masses inferred with the VLA data use a different dust opacity model and observe at longer wavelengths. Their light is thus more likely to be optically thin and may more accurately represent the real spread in dust masses in Class 0/I systems. The observed range of Class 0 dust masses span a similar mass range as we found for our modelled embedded disks when in their Class 0 phase. During the Class I phase, the modelled disks over-predict the dust mass compared to the observations. Both the data inferred by ALMA and the VLA conclude the same property of Class 0/I disks: that Class 0 disks tend to be more dust-rich than Class I disks.

Our model tends to predict that the embedded disk is more dust rich in its Class I phase than in its Class 0 phase, owing to the fact that it takes time for the dust from the envelope to fall into the disk. Furthermore, dust that reaches the disk early can already radially drift into the host star. This appears to be a major component of dust evolution in embedded disks since successive time steps in Figures \ref{fig:smdust_disk_buildup} and \ref{fig:smdust_em13} typically show that the disk is mainly populated by dust that has only recently arrived from further out in the envelope; dust that had arrived from earlier time steps are no longer found in the disk. This appears to be independent of the initial angular momentum in the cloud because both the reference and standard models share this property. Since our Class I disks are heavier than Class 0 they tend to overestimate the dust mass in observed Class I systems, except for the Std-lrg model where the large grains radially drift quickly in the embedded disk. Unless dust growth is very fast in the first few 10s of kyr into the collapse, it seems unlikely that the observed Class 0/I dusty disk masses are dominated by mm-dust grains because of their weak retention in the embedded disk.

\subsubsection{Impact of grain growth}

As argued in the introduction there is observational evidence to support a wide range of possible grain sizes in young stellar systems. Our results suggest that these populations of large grains are unlikely to be part of the initial cloud's dust grain distribution, since the bulk of the dust mass larger grains is rapidly lost to radial drift during the Class I phase. On the other hand, the smaller grains take between 0.3 - 0.45 $t_{\rm acc}$ ($\sim$75-113 kyr) to build up the observed median dust masses in the standard model. To account for the observational evidence of larger grains in proto-stellar systems, some amount of grain growth would need to occur during the collapse.

\ignore{
Our models have ignored the impact of grain growth when evolving the dust collapse. While this is reasonable in the falling envelope where growth timescale would be long, once the dust reaches the higher density embedded disk one might expect grain growth to occur. Indeed \cite{Drazkowska2018} compute the growth and evolution of dust during the Class I phase and found growth to be significant in the embedded disk. The growth would transfer mass from the smaller dust grains to larger, and potentially increase the efficiency that dust is lost in both the Class 0/I phases. This extra mass sink/source for the smallest/largest grains could explain why our medium and small dust grain size models over predict the mass observed in young proto-stellar systems, but may not help to explain why Class 0 embedded disks are heavier in dust mass than Class I disks. Computing the self-consistent growth of dust particles during the collapse and embedded disk build up is needed, and is currently beyond the scope of this work.
}

As previously mentioned their is a traffic jam of dust at the upper edge of the disk which could lead to some grain growth that would change the overall mass evolution of the disk. The typical growth timescale through coagulation is \citep{B12}:\begin{align}
    \tau_{\rm grow} \sim \frac{1}{\Omega\epsilon}.
\end{align}
As an estimation of the possible grain growth within the traffic jam we consider the 0.5 $t_{\rm acc}$ snapshot of the Std-med model (bottom-left panel of Figure \ref{fig:smdust_gas-to-dust}). Near 10 AU the thickness of the traffic jam region is approximately 1 AU, and the dust-to-gas ratio is within a factor of a few higher than the typical ISM ratio. The amount of time that a dust parcel moving at 100 cm s$^{-1}$ would take to cross the traffic jam region at 10 AU is almost 5 kyr. If the dust-to-gas ratio at 100 AU near the disk edge is equal to the typical ISM ratio then the growth timescale of the dust grains is at most 0.5 kyr. Since the disk is marginally more dust-rich than in the ISM (slightly more orange in Figure \ref{fig:smdust_gas-to-dust}) then the growth timescale is even smaller. Given that the growth timescale is smaller than the crossing time scale it is possible that some grain growth could occur in the traffic jam region.

These estimates are sensitive to the local dust-to-gas ratio and the kinematics of the grain parcels moving through the upper regions of the disk. This simple experiment, however, shows that grain growth near the upper edge of the embedded disk could have important implications on the overall evolution of the dust. In particular, grain growth would tend to produce larger grains which would contribute to the dust grain population simulated in the Std-lrg model. These larger grains can settle more quickly to the midplane, even in the presence of vertical turbulence and diffusion, and would provide a larger and more prolonged mass flux to the midplane than was found in the Std-lrg model. Given that the streaming instability is more efficient for larger grains, we would then expect more planetesimal mass to be produced. We leave the study of this possible effect to future work, particularly a simulation that self-consistency computes dust grain size evolution is crucial to include to the collapse.

Even with this possible region of the efficient grain growth, this does not fully explain the presence of large dust grains in the \textit{envelope}. For this, grain growth in the embedded disk would need a second mechanism to cycle the larger grains outward into the envelope where it could be observed. Such a mechanism would also help to sustain dust mass in proto-stellar systems, and is discussed below.

\subsubsection{Sustaining dust mass in Class 0/I disks}

In the context of our own Solar system, there have been models suggesting that dust grains are cycled from the inner disk outward to larger radii. Such a mechanism would counterbalance radial drift and could help to accelerate the dust mass build up during the Class 0 phase. These models centre around calcium-aluminium inclusion (CAI) formation and broadly argue that CAIs are formed in the high temperature region of the young Solar nebula before being mixed outward by an X-wind \citep{Shu1996,Shu1997,Shu2001}, MHD-driven winds and turbulence \citep{Cuzzi2003a,Cuzzi2003b}, or hydrodynamic turbulence \citep{Desch2010}. While these models generally focus on an older proto-stellar system than the $\sim$50 kyr Class 0 disk we discuss here - often including the presence of a young Jupiter for setting the distribution of CAIs \citep{McKeegan1998,Desch2018,vanKooten2021} - the physical mechanisms responsible for mixing CAIs throughout the young solar system could work against radial drift sufficiently to enhance the dust mass in Class 0 disks. The role of winds and the diffusive effect of turbulence on the radial evolution of the dust is not included in our model, but offer an interesting line of further study. 

In addition to helping understand the distribution of CAIs in the proto-solar nebula, the mixing of dust grains from the inner part of the proto-stellar envelope could also help to explain the existence of larger grains observed at large scales of young systems. While larger grains have been inferred observationally \citep{Galametz2019}, the physical mechanism to explain their existence in the outer envelope remains unsolved. If the cloud and its dust simply began collapsing from the surrounding ISM, the dust would be expected to be mainly monomer size. It is possible that these large grains are the result of early accretion of small grains into a Class 0 embedded disk, limited grain growth within the disk, followed by outward recycling back into the envelope. Solving this part of the problem may require large-scale hydrodynamic simulations to demonstrate whether this recycling is possible, and is beyond the scope of this work.

If an extra mechanism could be included that slowed the global impact of radial drift, then our overestimation of the Class I disk mass would be exasperated. However preserving dust mass in the disk, coupled with a certain amount of grain growth will lead to a higher available dust mass for the streaming instability. So while the preserved mass in Class I disks may be higher, the production rate of planetesimals would also be higher unless the dust mass is kept away from the midplane. This seems unlikely because as grains grow they will also tend to sink to the midplane - ready to be converted to planetesimals. If the physical mechanisms that are left out of our simple experiment tend to preserve dust mass then it is possible that the generation of the planetesimals can be efficient enough to reverse the tendency we see here - that Class I embedded disks contain more dust mass than Class 0 disks.

\section{Conclusion} \label{sec:conclude}

Here we model the collapse of a dusty proto-stellar envelope to study the accumulation of the dusty embedded disk. The variations in the collapse trajectory of dust grains, caused by hydrodynamic drag and the immediate collapse of dust can generate regions in the embedded disk where the local dust-to-gas ratio exceeds the typical 1\% of the ISM. These regions of enhanced dust mass become susceptible to the streaming instability which would lead to the generation of planetesimals very early in protostellar evolution. This process, and the subsequent beginning of planet formation is most efficient in the Class I phase for grain sizes $\leq 34$ $\mu$m, but even earlier in the Class 0 phase for larger grains. This generation of planetesimals can help to understand observations made in surveys of Class 0/I systems that show that Class I disks tend to be less massive than their younger Class 0 cousins. 

We have developed a simple model to account for the gas drag on a spherical cloud of dust embedded in the gas of a collapsing core. Our model uses the semi-analytic gas model of \cite{Visser2009} and computes the trajectory of 100,000 dust parcels as they collapse. This model includes the effect of gas drag on the dust particles as they fall towards the host star, which changes their collapse trajectory with respect to the gas particles. Overall we find that the dust grain's lack of gas pressure support causes them to fall much faster than the gas, forming dust enhancements along the midplane of the embedded disk. These enhancements survive for an extended period of time against radial drift along the midplane of the embedded disk because of ongoing mass flux from the collapsing envelope. This balance, however, tends to lean towards radial drift for the larger grains which are eventually lost to the host star on a timescale related to the grain size.

Through our simple experiment we have found that:\begin{enumerate}
    \item All grain sizes form regions of their embedded disk where planetesimals can be formed, with the larger grains doing so earlier.
    \item Our models have between 7 and 35 M$_\oplus$ available for the generation of planetesimals from the streaming instability during the Class 0/I phases.
    \item The sub-micron sized grains reach a total available mass of about 35 M$_\oplus$ near the transition between Class I and II. 
    \item The range of formed planetesimal masses is on the order of the shift in median masses between Class 0 and Class I objects as inferred by continuum emission surveys.
    \item Our total dust masses, with the mass available for the streaming instability removed, are larger than the median dust masses in observed Class I disks - representing the top $\sim$ 20\% of observed disk masses.
\end{enumerate}

The differences between our model and observations is likely linked to missing physics such as the outward radial mixing of dust as required to explain the distribution of CAIs in the Solar system.  Furthermore we have neglected turbulent diffusion, which can act to suppress mass clumping along the disk midplane (through the suppression of settling) and thus the streaming instability, but can also further counteract radial drift to slow the loss of dust mass. It is clear that while this semi-analytic approach to the collapse of a dusty pre-stellar cloud is illustrative, more rigorous numerical methods are needed to fully explore the mass evolution of young disks. 

The quantity of planetesimal mass that is generated in our models \textit{could} represent the first step in the planet formation process. Specifically, since the region of the embedded disk where planetesimals can be produced is restricted to roughly 10 AU, those planetesimal could hypothetically coagulate into the \textit{first} planetary core early on in the Class 0/I phase. This young planetary core (with mass a core mass of $\sim$20 M$_\oplus$) would then produce the first pressure bump in the embedded disk gas, causing dust trapping, and the earliest substructures in embedded disks. These dust traps would also promote further planetesimal growth as they are a prime location for high dust-to-gas ratios, which can further generate new planet cores. Thus, our model provides a minimal mass in planetesimals in this phase of star formation.

Under this scenario the formation of our own solar system would proceed with the formation of Jupiter's core first, producing a pressure maxima where the planetesimals which form Saturn's core are collected. Saturn's core would then grow and produce its own pressure maxima where the next generation of planetesimals would form. This next generation of planetesimals would proceed to form the ice giants. Saturn's delayed formation with respect to Jupiter would be in line with the assumptions of the Grand Tack model \citep{Walsh2011}, while the further delay of Uranus and Neptune's formation would explain their lower masses in the pebble accretion paradigm \citep[see for ex. ][]{Helled2020,Brouwers2021}.

It should be clear from our work that the generation of planetesimals through a process like the streaming instability can potentially occur during the collapse phase of proto-stellar formation. As such, this pushes the start time of the planet formation process back - possibly as far as into the Class 0 phase. Such an early start suggests that the first planets to form around a young proto-star may emerge nearly fully formed by the time their host star transitions from the Class I to the Class II phase - the colloquially called `proto'-planetary disk phase. 

\begin{acknowledgements}

 Astrochemistry in Leiden is supported by the Netherlands Research School for Astronomy (NOVA). B.T. acknowledges support from the research programme Dutch Astrochemistry Network II with project number 614.001.751, which is financed by the Dutch Research Council (NWO). GR acknowledges support from the NWO (program number 016.Veni.192.233) and from an STFC Ernest Rutherford Fellowship (grant number ST/T003855/1).

\end{acknowledgements}

\bibliographystyle{aa} 
\bibliography{mybib.bib} 

\appendix
\section{ Collapse of the Std-lrg and Std-sml models }

\begin{figure}
    \centering
    \begin{minipage}{0.49\textwidth}
            \includegraphics[width=\textwidth]{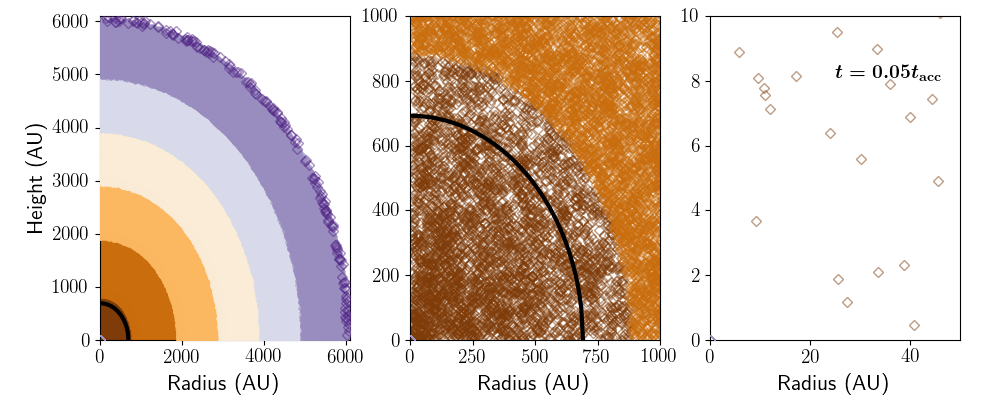}
    \end{minipage} 
    \begin{minipage}{0.49\textwidth}
            \includegraphics[width=\textwidth]{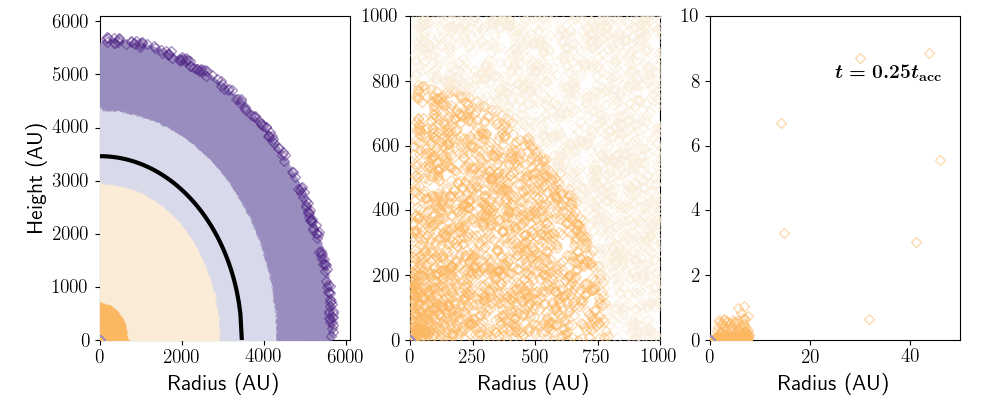}
    \end{minipage}
    \begin{minipage}{0.49\textwidth}
            \includegraphics[width=\textwidth]{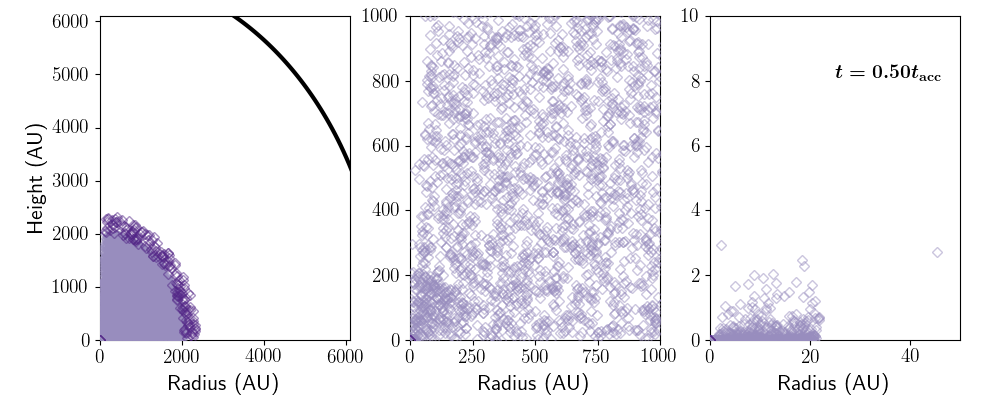}
    \end{minipage} 
    \begin{minipage}{0.49\textwidth}
            \includegraphics[width=\textwidth]{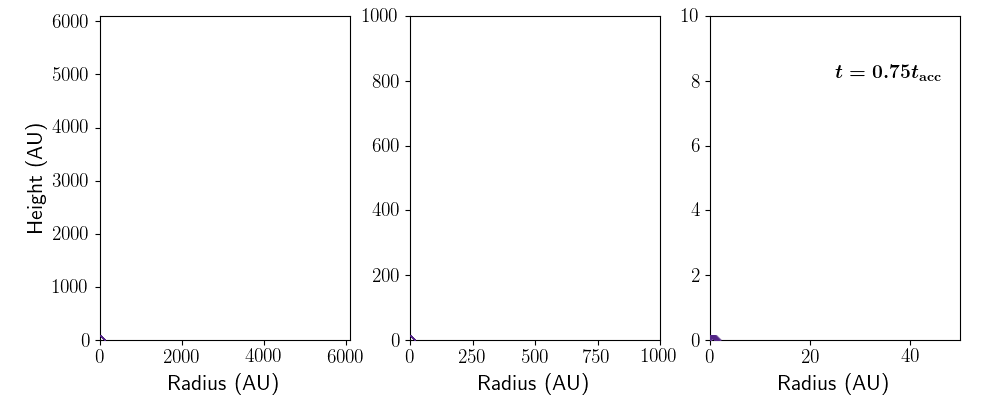}
    \end{minipage}
    \caption{ Same as in Figure \ref{fig:smdust_disk_buildup} but for the Std-lrg model. The larger grains are less coupled to the gas and collapse much more quickly. Once in the disk their radial drift is also faster than in the Std-med model, and they rapidly deplete into the host star. As such, unlike in the Std-med model grains originating from different initial radii ranges do not coexist in the embedded disk with dust parcels of different colours. }
    \label{fig:dust_disk_buildup}
\end{figure}

\begin{figure}
    \centering
    \begin{minipage}{0.49\textwidth}
            \includegraphics[width=\textwidth]{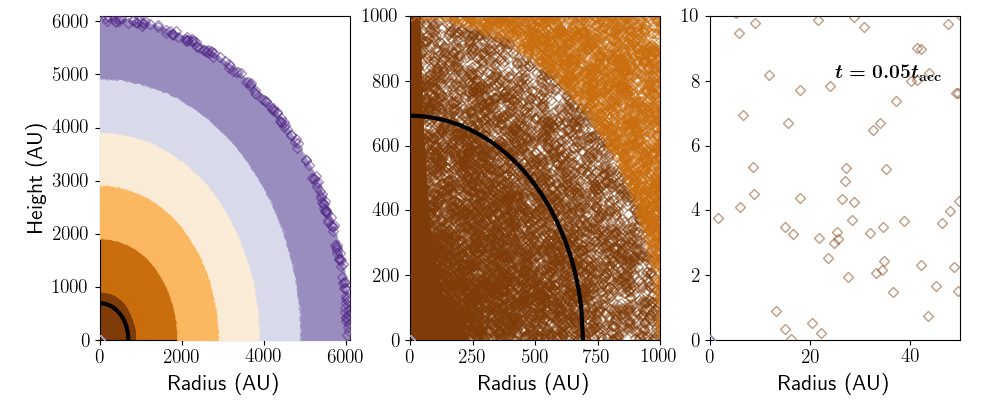}
    \end{minipage} 
    \begin{minipage}{0.49\textwidth}
            \includegraphics[width=\textwidth]{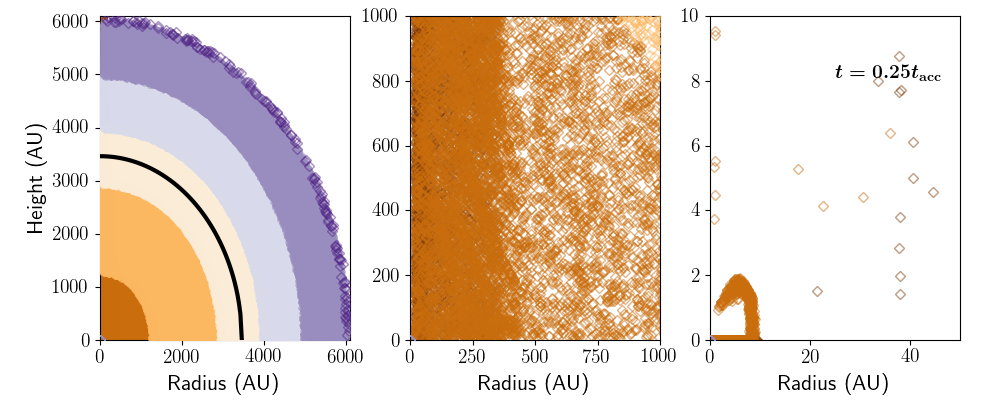}
    \end{minipage}
    \begin{minipage}{0.49\textwidth}
            \includegraphics[width=\textwidth]{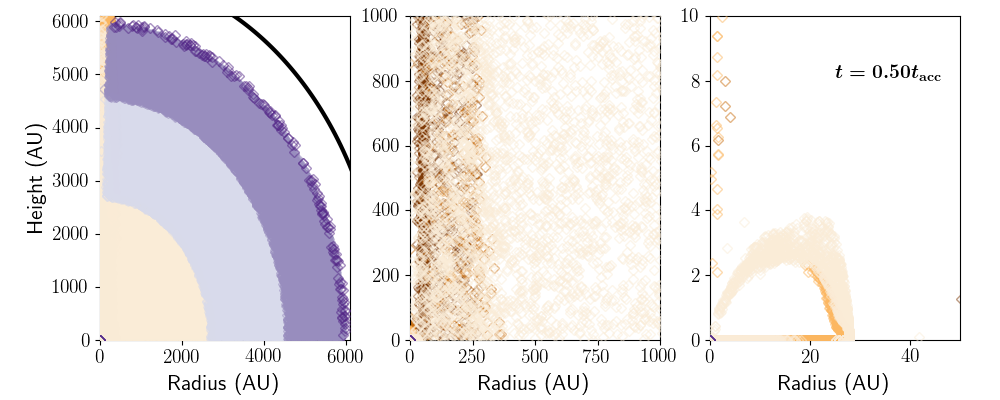}
    \end{minipage} 
    \begin{minipage}{0.49\textwidth}
            \includegraphics[width=\textwidth]{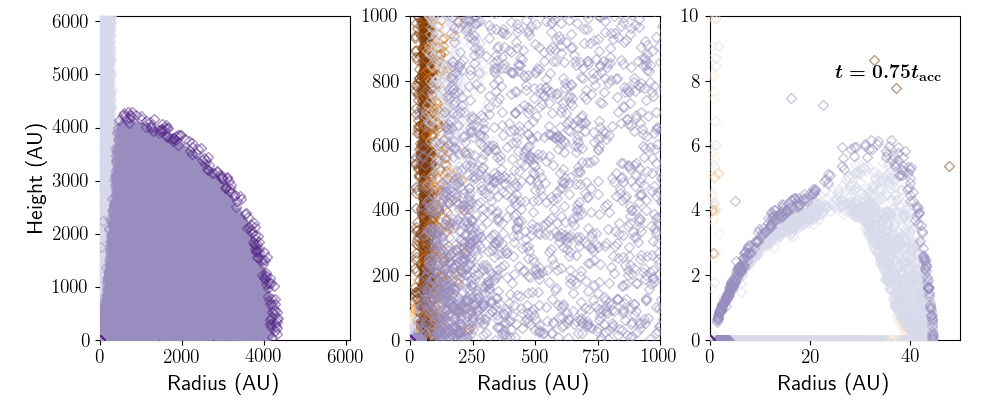}
    \end{minipage}
    \caption{ Same as in Figure \ref{fig:smdust_disk_buildup} but for the Std-sml model. These smaller grains are more dynamically coupled to the gas, and follow similar trajectories. The dust disk that they form are also more similar to the gas as radial drift and settling are both less efficient for this population of dust grain. }
    \label{fig:vsmdust_disk_buildup}
\end{figure}

For brevity in the main text we leave the collapse trajectories of the Std-lrg and Std-sml here. The colour of the points have the same definition as they did in Figures \ref{fig:gas_buildup}-\ref{fig:smdust_em13}. The grains in the Std-lrg model (figure \ref{fig:dust_disk_buildup}) evolve much faster than the small grains in the Std-sml model (figure \ref{fig:vsmdust_disk_buildup}) as can be seen by comparing the colour of the dust parcels at any given time.

\section{Gas-to-dust ratio in Std-lrg and Std-sml models}

\begin{figure*}
    \centering
    \begin{minipage}{0.49\textwidth}
            \includegraphics[width=\textwidth]{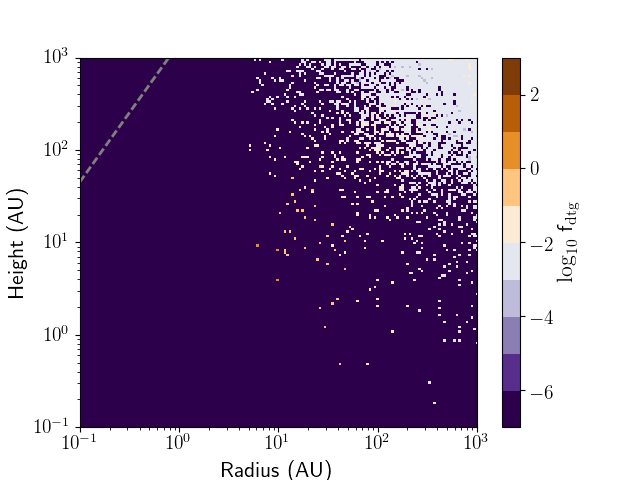}
    \end{minipage} 
    \begin{minipage}{0.49\textwidth}
            \includegraphics[width=\textwidth]{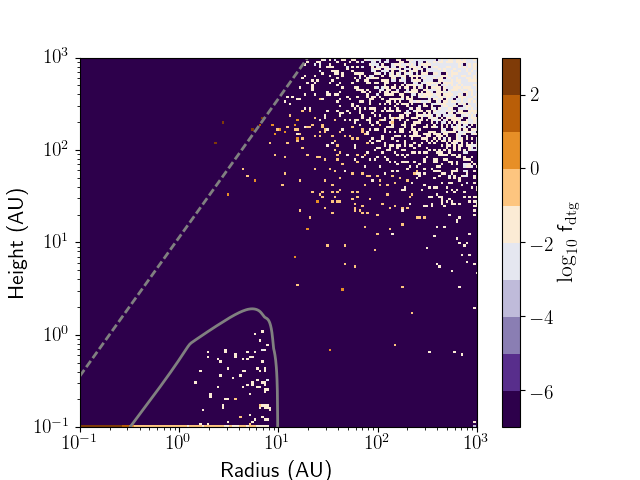}
    \end{minipage}
    \begin{minipage}{0.49\textwidth}
            \includegraphics[width=\textwidth]{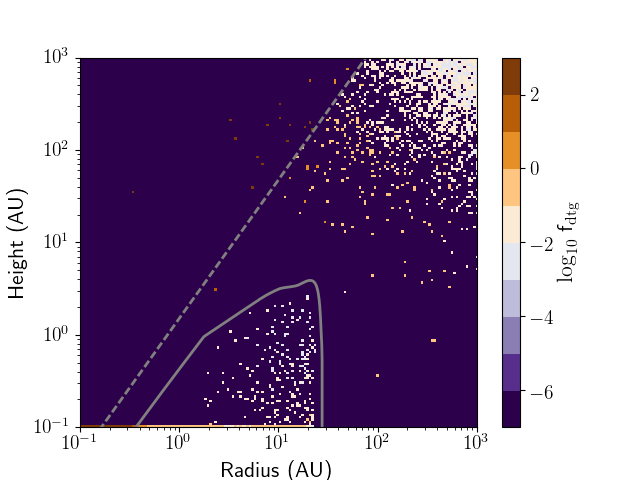}
    \end{minipage} 
    \begin{minipage}{0.49\textwidth}
            \includegraphics[width=\textwidth]{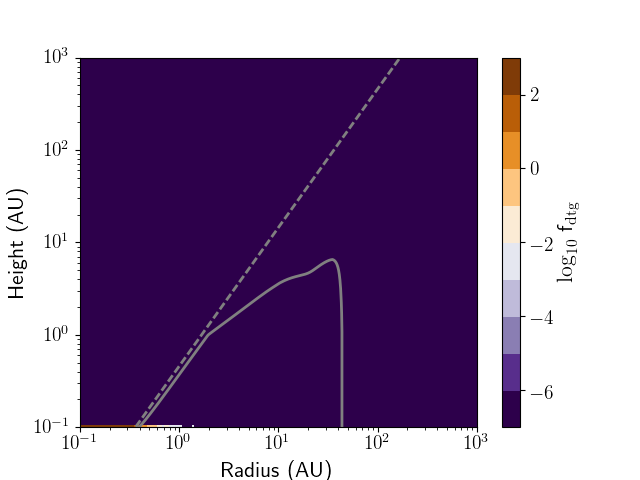}
    \end{minipage}
    \caption{ Same as Figure \ref{fig:smdust_gas-to-dust} but for the Std-lrg model. These larger grains evolve faster than smaller once, particularly in the high density of the embedded disk, they radially drift very efficiently into the host star. }
    \label{fig:dust_gas-to-dust}
    \vspace{5pt}
    \begin{minipage}{\textwidth}
    \begin{overpic}[width=0.9\textwidth]{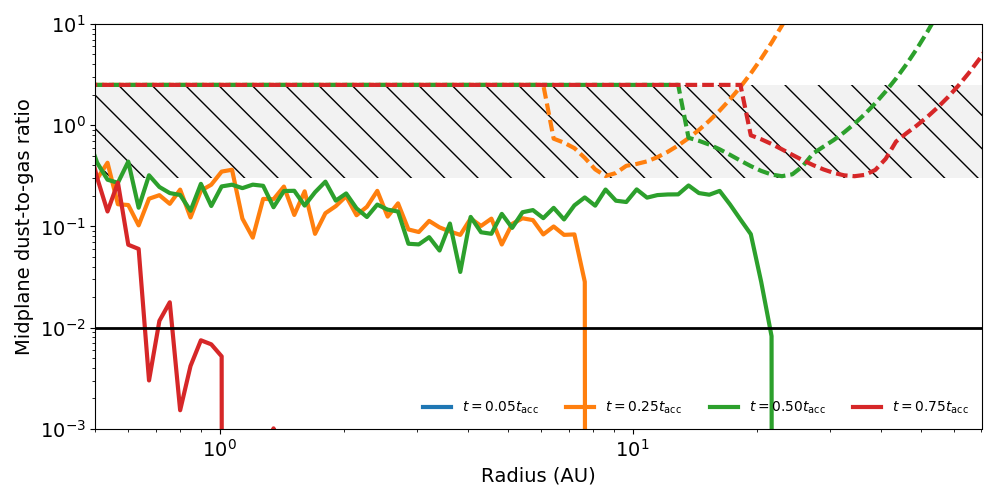}
        \put(50,18){\color{black}\Large ISM}
        \put(15,39){\color{black}\Large Threshold for streaming instability}
    \end{overpic}
    \caption{The midplane dust-to-gas ratios for the Std-lrg collapse model at the same time steps as shown in Figure \ref{fig:dust_gas-to-dust}. The lines and hashed regions have the same definition as in \ref{fig:midplane_smdust}. }
    \label{fig:midplane_dust}
    \end{minipage}
\end{figure*}

\begin{figure*}
    \centering
    \begin{minipage}{0.49\textwidth}
            \includegraphics[width=\textwidth]{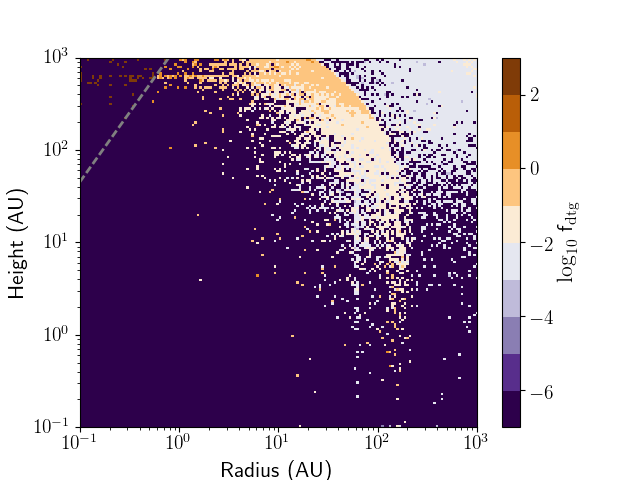}
    \end{minipage} 
    \begin{minipage}{0.49\textwidth}
            \includegraphics[width=\textwidth]{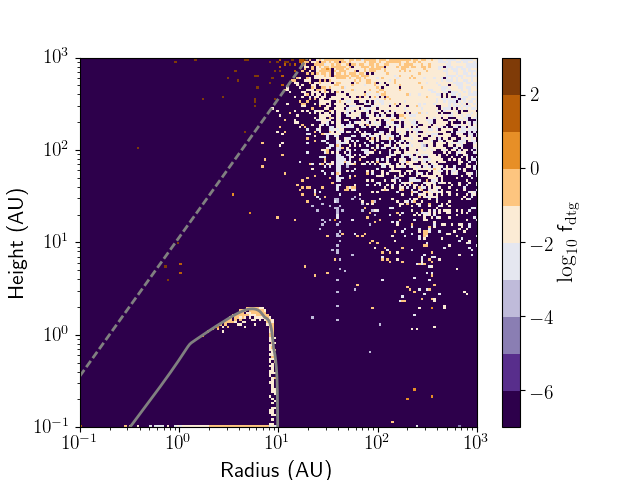}
    \end{minipage}
    \begin{minipage}{0.49\textwidth}
            \includegraphics[width=\textwidth]{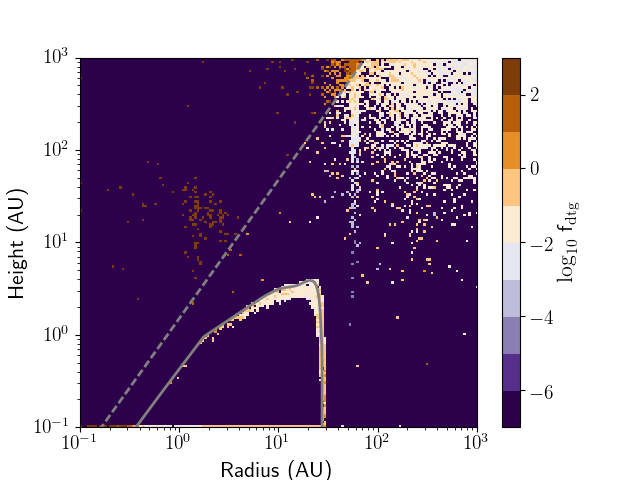}
    \end{minipage} 
    \begin{minipage}{0.49\textwidth}
            \includegraphics[width=\textwidth]{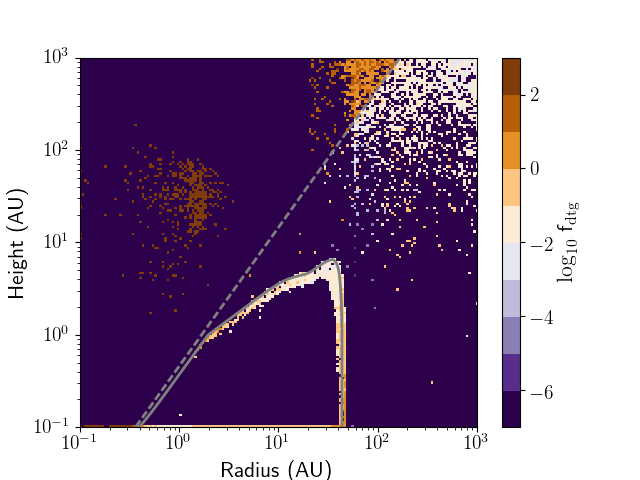}
    \end{minipage}
    \caption{ Same as Figure \ref{fig:smdust_gas-to-dust} but for the Std-sml model. Because of the small dust grains used in this model the collapse and evolution in the disk is more coupled to the gas. As such they tend to evolve in a similar was as the gas, creating mainly ISM dust-to-gas ratios along the midplane of the disk. There are thin regions around the surface of the disk where the sudden pressure change captures the dust - similarly to typical dust traps in Class II disks.}
    \label{fig:vsmdust_gas-to-dust}
    \vspace{5pt}
    \begin{minipage}{\textwidth}
    \begin{overpic}[width=0.9\textwidth]{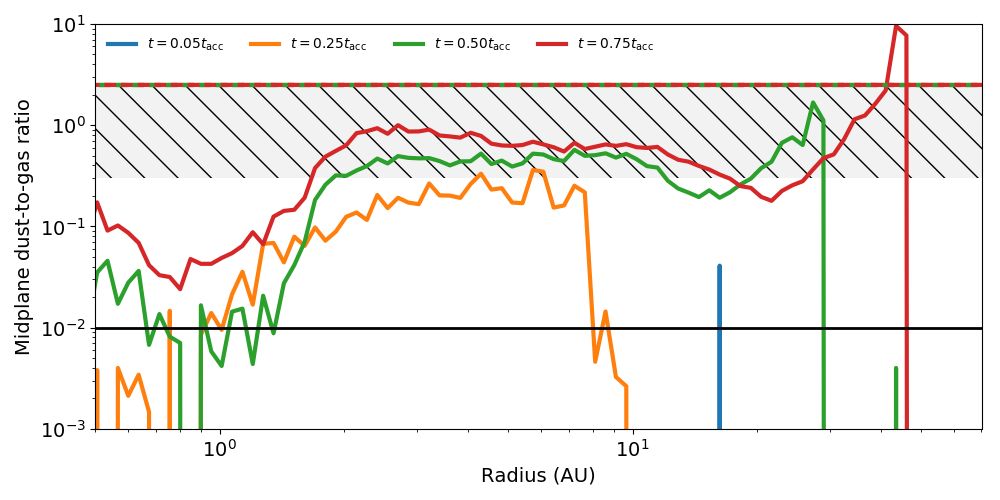}
        \put(50,18){\color{black}\Large ISM}
        \put(15,39){\color{black}\Large Threshold for streaming instability}
    \end{overpic}
    \caption{The midplane dust-to-gas ratios for the Std-lrg collapse model at the same time steps as shown in Figure \ref{fig:vsmdust_gas-to-dust}. The lines and hashed regions have the same definition as in \ref{fig:midplane_smdust}.}
    \label{fig:midplane_vsmdust}
    \end{minipage}    
\end{figure*}

Again we place the results of the Std-lrg and Std-sml here for brevity in the main text. As discussed it is clear from Figures \ref{fig:midplane_dust} and \ref{fig:midplane_vsmdust} the dust in the Std-lrg model rapidly evolve, but still produce regions of the dist that exceeds a dust-to-gas ratio of $\sim 0.5$ which is enough to form planetesimals. Meanwhile the dust in the Std-sml dust model takes longer to build up along the disk midplane, but due to its slow radial drift in the disk, the dust-to-gas ratio can rapidly build up at later times, producing exceedingly large dust-to-gas ratio near the end of the simulation.

\end{document}